\documentclass[useAMS, usenatbib]{mn2e}
\usepackage{graphicx,lscape, subfigure}
\usepackage{float}
\usepackage{color}
\usepackage{ram}
\usepackage{times}
\usepackage[T1]{fontenc}
\usepackage{aecompl}

\newcommand{\ant}{\rm {\alpha_{nt}}}
\begin{document}

\title[GMRT radio continuum study of Wolf Rayet galaxies I: NGC 4214 and NGC 4449]{GMRT radio continuum study
of Wolf Rayet galaxies I:NGC 4214 and NGC 4449}
\author[Srivastava et al.]{Shweta Srivastava$^{1}$\thanks{E-mail:srivashweta@gmail.com},
N. G. Kantharia$^{2}$,
Aritra Basu$^{2}$,
D. C. Srivastava$^{1}$,
S. Ananthakrishnan$^{3}$\\
$^{1}$Dept. of Physics, DDU Gorakhpur University, Gorakhpur - 273009, India\\
$^{2}$National Centre for Radio Astrophysics, TIFR, Pune - 411007, India\\
$^{3}$Dept. of Electronic Science, Pune University, Pune - 411007, India}

\maketitle

\begin{abstract}

We report low frequency observations of Wolf-Rayet galaxies, NGC 4214 and NGC
4449 at 610, 325 and 150 MHz, using the Giant Meterwave Radio Telescope (GMRT).
We detect diffuse extended emission from NGC 4214 at and NGC 4449.  NGC 4449 is
observed to be five times more radio luminous than NGC 4214, indicating
vigorous star formation. We estimate synchrotron spectral index after separating the thermal
free-free emission and obtain $\alpha_{nt}=-0.63\pm0.04$
(S$\propto\nu^{\alpha_{nt}}$) for NGC 4214 and $-0.49\pm0.02$ for NGC 4449.
About $22\%$ of the total radio emission from NGC 4214 and $\sim 9\%$ from NGC
4449 at 610 MHz is thermal in origin.  We also study the spectra of two compact
star-forming regions in NGC 4214 from 325 MHz to 15 GHz and obtain
$\alpha_{nt}=-0.32\pm0.02$ for NGC 4214-I and $\alpha_{nt}=-0.94\pm0.12$ for
NGC 4214-II.  The luminosities of these star-forming regions ($\sim 10^{19}\rm
W~ Hz^{-1}$) appear to be similar to those in circumnuclear rings in normal
disk galaxies observed with similar linear resolution.  We detect the supernova
remnant SNR J1228+441 in NGC 4449 and estimate the spectral index of the
emission between 325 and 610 MHz to be $-1.8$ in the epoch 2008-2009.  The
galaxies follow the radio-FIR correlation slopes suggesting that star formation
in Wolf - Rayet galaxies, which are low-metallicity systems, are similar to
that of normal disk galaxies.

\end{abstract}


\section{Introduction}

Wolf - Rayet (WR) galaxies show broad emission lines in the optical spectrum
due to WR stars. Broadband studies of the spectrum, from radio to optical, of
the dwarf galaxy He 2-10 indicated the presence of $10^3$ to $10^4$ WR stars
\citep{allen1976}.  The discovery of WR galaxies has been serendipitous until a
catalogue of 37 WR galaxies was first compiled by \citet{conti1991}. 139 new
members were added by \citet{schaerer1999}.  Currently 846 WR galaxies are
known, following the SDSS Data Release 6 \citep[DR6;][]{adelman2008}.  WR
features in a galaxy provide useful information about the star-formation
processes in the system like the age and strength of the starburst. The WR
phase is characterised by the ejection of the outer layers of evolved massive
stars by stellar winds.  These galaxies, along with blue compact dwarf and
irregular star-forming galaxies, are collectively referred to as H{\sc ii}
galaxies. 

\begin{table*}
\caption{General Properties of the Galaxies}
\begin{tabular}{@{}llllllllll@{}}
\hline
Galaxy   &  Right Ascension$^{(1)}$ & Declination$^{(1)}$ & Distance & $d_{25}$$^{(5)}$ & $m_B$$^{(5)}$ & 12+log$\frac{O}{H}$ & $v_{hel}$$^{(10)}$ & F$_{H\alpha}$ \\
         &  (J2000)         & (J2000)     & (Mpc)      &  (arc min)  & (mag)  &  & $({\rm km s^{-1}})$ &$({\rm ergsec^{-1}cm^{-2}})$ \\  
\hline\hline
NGC 4214 & $12^h15^m39.2^s$ & $+36^{\circ}19'37''.0$ & 2.94$^{(2)}$ & $8.91\pm1.32$& $10.17\pm0.21$ & $8.28\pm0.08$$^{(6)}$& $291\pm3$ & $1.47\times10^{-11}$$^{(8)}$\\
NGC 4449 & $12^h28^m11.9^s$ & $+44^{\circ}05'40''$& 3.9$^{(3)}$ & $5.63\pm0.29$& $9.65\pm0.62$& $8.31\pm0.07$$^{(7)}$& $202\pm7$& $2.03\times10^{-11}$$^{(9)}$\\
\hline
\end{tabular}
\\
\label{tab1:Galaxy Parameters}
\footnotesize{References: (1) NED; (2) \citet{maiz2002}; (3) \citet{annibali2008}; 
(4) \citet{deVaucouleurs1991}; (5) \citet{paturel1997}; (6) \citet{kobulnicky1996};
(7) \citet{martin1997}; (8) \citet{martin1998}; (9) \citet{hunter1993}; (10) \citet{bottinelli1990}  }\\
\end{table*}

Some of these H{\sc ii} galaxies have been studied at radio wavelengths.  A
sample of 26 blue compact dwarf (BCD) galaxies was studied at high radio
frequencies ($>$1 GHz) by \citet{klein1991}.  For five galaxies, they found a
steep non-thermal spectrum ($\alpha < -0.7$) after separation of the thermal
free--free component of the radio emission.  They suggested a new relation
between the radio spectral index and optical luminosity where they find that
the spectral index is steeper for luminous objects.  The galaxies were found to
follow the radio - FIR correlation observed in normal disk galaxies.  A high
resolution radio continuum study between 1.4 GHz and 15 GHz along with
H$\alpha$ observations of nine WR galaxies revealed flat ($\alpha > -0.4$)
radio spectra and in many cases inverted spectra between 5 and 15 GHz
\citep{beck2000}.  They explain that this is due to WR galaxies harbouring
young starbursts and hence the spectrum is dominated by free-free emission from
the star-forming regions. However, a spectral study of seven such H{\sc ii}
galaxies at 325 MHz by \citet{deeg1993} revealed a range of spectral behaviour.
They noted that a variety of emission, absorption and energy-loss mechanisms
are responsible for the shape of the radio spectra.  They do not find steep
non-thermal spectra in their sample, unlike \citet{klein1991}. A recent radio
continuum study of five BCD galaxies using the Giant Metrewave Radio Telescope
(GMRT) by \citet{ramya2011} also resulted in a range of spectral shapes at low
radio frequencies, which they suggest could be indicative of environmental
effects.  They found that the extended radio emission is generally detected in
galaxies residing in a group environment whereas localised emission, which
seems to correlate with the H$\alpha$ emission, is detected in galaxies
evolving in a relatively isolated environment.  Thus, the former shows a
steeper spectrum as compared to the latter.  However, their sample is small and
it will be interesting to check for a larger number of H{\sc ii} galaxies.  

It is important to increase the sample of H{\sc ii} galaxies studied at low
radio frequencies and disentangle the different physical mechanisms which play
a role in shaping their evolution and spectra. We are studying the low radio
frequency properties of a sample of seven WR galaxies using the GMRT.  We plan
to combine this with the higher radio frequency data and model their spectra,
separate the thermal contribution to the spectrum and estimate the non -
thermal spectral index.  It will be interesting to examine the influence of
environment on the morphology of the radio continuum emission and the low radio
frequency spectrum and compare the low radio frequency continuum properties of
these low-metallicity systems with disk star-forming galaxies.

In this paper, we present low-frequency study of two well-known WR galaxies of
our sample, namely, NGC 4214 and NGC 4449.  Both galaxies are classified as
Magellanic irregular and are located in the Canes Venatici region. Some
properties of these galaxies are listed in Table \ref{tab1:Galaxy Parameters}.
In \textsection 2 we describe observations, data calibration and imaging
processes.  The results are presented in \textsection 3 and discussed in
\textsection 4. The conclusion is presented in the last section.  We use the
convention of spectral index $\alpha$ defined via $S\propto\nu^\alpha$.  

\subsection{NGC 4214}

NGC 4214 is a nearby barred S-shaped Magellanic irregular galaxy having nearly
face-on orientation \citep{allsopp1979}.  It has been identified as a starburst
galaxy with an inner starburst region and an older red disk and has been the
subject of extensive multi-wavelength research. NGC 4214 has two main regions
of star formation - the northern one (RA=12h15m39s, DEC=$36^{\circ}19'35''$,
J2000) has a shell-type morphology visible at high angular resolution near the
centre of the galaxy which is known in literature as NGC 4214-I.  A smaller
second star forming complex located to the south (RA=12h15m40s,
DEC=$36^{\circ}19'09''$, J2000) of NGC 4214-I is known as NGC 4214-II.  NGC
4214 is a member of the group LGG 291 \citep{garcia1993} which contains 14
galaxies.   The two closest members are the dwarf galaxy DDO 113 and the
star-forming irregular galaxy NGC 4190 having similar radial velocities as NGC
4214. These galaxies are within angular separation of $\leq 30'$ from NGC 4214
and consequently hace been in the field of view. The distance to NGC 4214
reported in literature ranges from 2 Mpc to 7 Mpc.  Throughout this paper, we
use a distance of 2.94 Mpc  \citep{maiz2002} and thus 1$''$ corresponds to 14
pc.  NGC 4214 is a gas-rich blue galaxy with a thick disk \citep{maiz1997}.

Diffuse X-ray emission from hot gas is detected in this galaxy
\citep*{ott2005}. Intense UV emission observed from this galaxy is believed to
be due to its low dust content, leading to inefficient processing of UV to
longer wavelengths \citep{fanelli1997}.  \citet{mackenty2000} detected several
young ($<$10 Myr) star-forming complexes of various ionised gas morphologies
with sizes $\sim$10-200 pc in their HST observations.  In addition, they
reported that the extended diffuse ionised gas in NGC 4214 contributes 40$\%$
of the total H$\alpha$ and O III $\lambda$5007$\AA$ emission of the galaxy.
The H{\sc i} gas in this galaxy is observed to be $\sim1.4$ times the Holmberg
radius in extent \citep{allsopp1979}.  
NGC 4214 has been studied in the radio continuum by \citet{allsopp1979,
beck2000, mackenty2000} and recently by \citet{kepley2011}.  \citet{beck2000}
studied the radio continuum emission from the star forming regions NGC 4214-I
and NGC 4214-II at 15, 8.3, 5 and 1.4 GHz using VLA at high angular resolution.
Interestingly, they found that the spectrum was rising between 5 and 15 GHz,
indicating the presence of optically thick free-free thermal emission.
\citet{kepley2011} estimated a magnetic field of $30 \mu$G in the central parts
and $10 \mu$G at the edges and do not detect significant polarisation on scales
larger than 200 pc. 

\subsection{NGC 4449}

NGC 4449 is a typical type 1 Magellanic irregular galaxy with star formation
occurring throughout the galaxy at a rate almost twice that of the LMC
\citep{thronson1987, hunter1999}.  The distance to the galaxy reported in the
literature ranges from 2.9 Mpc to 5.4 Mpc.  Throughout this paper, we use a
distance of 3.9 Mpc given by \citet{annibali2008} and thus $1''$ corresponds to
19 pc.  This galaxy has been studied extensively in the optical bands ( e.g.
\citet{crillon1969, kewley2002}), in the radio \citep{seaquist1978, klein1986,
hunter1991, klein1996}, in the H{\sc i} \citep{rogstad1967, vanWoerden1975} and
in  CO \citep{tacconi1985, sasaki1990, bottner2003}.  NGC 4449 appears to be
undergoing tidal interaction with the neighbouring dwarf galaxy DDO 125
\citep{hunter1998, theis2001}.  

\citet{klein1996} have studied NGC 4449 using WSRT and detected a large
synchrotron halo ($\sim 7$ kpc) around the central star forming region at 610
MHz.  At the distance of 3.9 Mpc, the halo extent is about 7.4 kpc
\citep{klein1996}.  They found the radio spectrum of the synchrotron radiation
between 610 MHz and 24 GHz to be $\sim -0.5$ in the central regions and $\sim
-0.7$ in the halo regions.  They fitted a combination of thermal and
non-thermal emission to the integrated spectrum between 610 MHz and 24 GHz and
found the global non-thermal spectral index to be $-0.7$.  This galaxy is found
to be rich in H{\sc i}, with M$_{HI}$ = $2.4\times10^{9}{\rm M_\odot}$
\citep{epstein1964} and the H{\sc i} halo extends to $\sim$6 times the Holmberg
radius \citep{hunter1998}.
X-ray observations of NGC 4449 show a circular region of diffuse emission
$\sim$ 1$'$ north of the centre of the galaxy \citep{ott2005}.  
\citet{chyzy2000} studied the polarised radio continuum emission of NGC 4449,
at 8.46 and 4.86 GHz and found a regular galaxy-scale  magnetic field of
strength $6-8 \mu$G.  They point out that the strong magnetic fields are
intriguing since the galaxy exhibits slow rotation which cannot support the
dynamo action required to amplify the magnetic field.  They also detect the
large continuum halo at 8 GHz.

\section[]{Observations and Data Reduction}

Both the galaxies were observed using the GMRT; \citep{swarup1991, ananth2005}
in the radio continuum at 150, 240, 325 and 610 MHz.  The 240 and 610 MHz
observations were simultaneous in the dual band mode.  It is remarked that the
240 MHz data of both the galaxies were corrupted to the extent that even after
extensive flagging the final image quality was suspect. We hence did not use
240 MHz data in the analysis presented here.  Since the galaxies are closely
located in the sky, they were observed in the same observing run with common
calibration scans.  The target source scans alternated between the two galaxies
resulting in better {\it uv-}coverage for both the galaxies.    

The 240 and 610 MHz observations were simultaneous in the dual band mode.  All
the observations were done in the non-polar mode of observing known as the
Indian polar mode. The details of these observations are given in
Table~\ref{tab2:Observational Parameters}.  We also analysed GMRT archival data
at 610 MHz for NGC 4214 from August 2007.  These single frequency observations
have better sampling of short spacings and hence are presented in this paper.  

\begin{table*}
\centering
\caption{Our Observations}
\label{tab2:Observational Parameters}
\begin{tabular}{@{}llll@{}}
\hline
Observing Band (MHz)                & 150          &  325          &  610$^{\#}$           \\
\hline\hline
Observation date                 & 28/10/2012 & 23/12/2009 & 20/12/2008 \\
Pointing Centre - NGC 4214  & $\alpha = 12^h15^m39.2^s$ and $\delta = +36^{\circ}19'37''.0$ & & \\
Pointing Centre - NGC 4449  & $\alpha = 12^h28^m11.9^s$ and $\delta = +44^{\circ}05'40''.0$ & & \\
Receiver Bandwidth (MHz)            & 16             &  32           &  16            \\
No. of Antennas$^{*}$                & 27             &  29           &  28             \\
Primary Calibrator(s)               & 3C286             &  3C147        &  3C147, 3C286  \\
Flux density (Jy)$^{\dagger}$       & 31.05             &  52.69        &  38.12, 21.01   \\
Secondary calibrator(s)             & 3C286             & 1123$+$055     &  1227$+$365         \\
Flux density (Jy)$^{\ddagger}$      & 31.05             & 6.7$\pm$0.15  &  1.8$\pm$0.03             \\
Telescope time (hrs)                & 9             & 7         &  7              \\
\hline
\end{tabular}
\\
\footnotesize{$^{\#}$ Dual band observation (240 \& 610 MHz). 
$^*$ Maximum number of operational antennas  during the observation. 
$\dagger$ Set by SETJY task : Using (1999.2) VLA or Reynolds (1934-638) coefficients. 
$\ddagger$ Flux density and error from GETJY}\\
\end{table*}

\begin{table*}
\caption{Image parameters}
\label{tab3:Achieved Parameters}
\begin{tabular}{@{}lllllll@{}}
\hline
Galaxy Name              & \multicolumn{3}{c}{NGC 4214}           & \multicolumn{3}{c}{NGC 4449}    \\
\hline
 Observing Bands    & 150 (MHz) &  325 (MHz) &  610 (MHz)      & 150 (MHz)      &  325 (MHz)         &  610 (MHz)       \\
\hline\hline
Beam Size(ROBUST=0) & $21''.7\times17''.8$ & $10''.5\times8''.6$   &  $5''.4\times4''.8$ & $20''.5\times17''.6$ & $11''.3\times8''.7$  & $5''.3\times5''.0$ \\
Position Angle & $71^{\circ}.7$ & $61^{\circ}.2$ & $47^{\circ}.2$ & $62^{\circ}.9$ &  $56^{\circ}.7$ &  $0^{\circ}.5$     \\
RMS noise level (mJy beam$^{-1}$) & 3.2          &  0.34             &  0.09             & 2.3                &  1.4               &  1.1    \\       \\
\hline
Beam Size (ROBUST=5)& $30''.3\times24''.4$ & $32''.0\times13''.3$ & $8''.5\times6''.5$ & $28''.8\times23''.6$ & $15''.3\times11''.1$ & $7''.6\times7''.1$ \\
Position Angle & $77^{\circ}.0$ & $3^{\circ}.3$ & $42^{\circ}.6$ & $65^{\circ}.5$ & $59^{\circ}.1$ & $5^{\circ}.2$ \\
RMS noise level (mJy beam$^{-1}$)  & 6.2             & 0.56             & 0.13             & 4.3              & 1.9              & 0.22     \\
\hline
Shortest spacing (k$\lambda$)     & 0.04             & 0.06             & 0.20     &0.04 &  0.06&  0.20      \\
                                  & $\sim 77$ m      & $ \sim55$ m      & $\sim 98$ m   &$\sim 77$ m &$ \sim55$ m& $\sim 98$ m       \\
Longest spacing (k$\lambda$)      & 13               & 27               &  50         &13 &27& 50    \\
                                  & $\sim 25$ km     & $\sim 25$ km     & $\sim 24$ km     & $\sim 25$ km & $\sim 25$ km &$\sim 24$ km    \\
Largest Visible structure$^{\#}$  & $\sim 52\arcmin$ & $\sim 34\arcmin$ & $\sim 10\arcmin$ & $\sim 52\arcmin$ & $\sim34\arcmin$ & $\sim10\arcmin$\\
\hline
\end{tabular}
\\
\footnotesize{Note: $^{\#}$ Corresponding to the shortest spacing present in our data.} 
\end{table*}

The observations started and ended with a 15m observing run on an
amplitude/bandpass calibrator (3C 286 or 3C 147).  The phase calibrator was
observed for about 5m after every 20m on a target galaxy. All the
observations except the last one in 2012 used the old, now decommissioned GMRT
hardware correlator.  The 2012 run used the new GMRT software backend. These
observations used a baseband width of 16 MHz.  The raw data obtained in the
local $lta$ format were converted into FITS and this data file was then
imported into the NRAO AIPS{\footnote{AIPS software is produced and maintained
by the National Radio Astronomy Observatory, a facility of the National Science
Foundation operated under cooperative agreement by Associated Universities,
Inc.}} software.  The data were edited, calibrated and imaged using standard
tasks in AIPS.  After an initial round of flagging, the single channel data of
the amplitude calibrator were calibrated and then the editing of bad data and
calibration were iteratively repeated.  The data for antennas with relatively
large errors in antenna-based gain solutions were examined and flagged.  Once
the gain calibration was satisfactory, the calibrated data were used to obtain
the bandpass calibration tables.  Once the bandpass calibration became
satisfactory, i.e. the closure errors were less than one percent, the bandpass
gain tables were applied to the data and the channel data averaged to generate
continuum data sets. To avoid bandwidth smearing at low frequencies due to
large fields of view, we averaged 20 channels at 610 MHz, 7 channels at 325 MHz
and 4 channels at 150 MHz corresponding to channel widths of 1.8 MHz, 0.9 MHz
and 0.4 MHz, respectively.  Thus a SPLAT data base with these channel widths
was generated for further analysis.  The amplitude and phase calibrator data
were used to obtain the gains of the antennas which were then applied to the
target source.   Bad data on the target source were edited and the final
calibrated visibility data were obtained.

These data were then imaged by creating multiple facets across the primary beam
to avoid wide-field effects in the image plane.  25 facets were used for
imaging data at 610 MHz whereas 49 facets were used at 325 and 150 MHz.  This
image was used as the model for the first round of phase-only self-calibration.
On average two to three rounds of phase-only self-calibration were required,
followed by a final round of amplitude and phase self-calibration.   This
procedure improved the images at all the frequencies.  All facets were combined
using the task FLATN to produce a final map at each observing frequency.
Moreover, we made maps using a range of different weighting schemes with
robustness parameters, where +5 is a pure natural weight and -5 is a pure
uniform weight \citep{briggs1995}.  This resulted in different beam sizes and
signal-to-noise ratios.  We also made images after including a uv cutoff in the
data to examine the presence of features on different angular scales.  The
final images were corrected for primary-beam attenuation before undertaking
flux density measurements. 

The presence of a strong source close to NGC 4449 affected the image quality
and to improve the image, we removed this source using the task UVSUB in AIPS.
While this did improve the images, it was difficult to remove all the
artifacts. In particular the final image at 325 MHz was badly affected and
therefore, we do not present the image but include the flux density estimates
at this frequency, with the caveat that the value quoted might have a large
systematic error.  

\section{Results}

The final images made at 610, 325 and 150 MHz for the two galaxies are shown in
Figures~\ref{fig3:4214.ROB0} and ~\ref{fig15:4449} and the image parameters
along with the results are listed in Table~\ref{tab3:Achieved Parameters}.
The flux densities for both the galaxies estimated at the different GMRT
frequencies are listed in Table~\ref{tab7:Flux Densities}, along with the
values obtained from literature.  The spectrum is shown in
Figure~\ref{fignew10:4449}.  It is to be noted that the systematic errors on
the GMRT flux denstities estimated using the noise statistics in a similar
region around the source were larger than the statistical errors. 

\subsection{NGC 4214}

\begin{figure*}
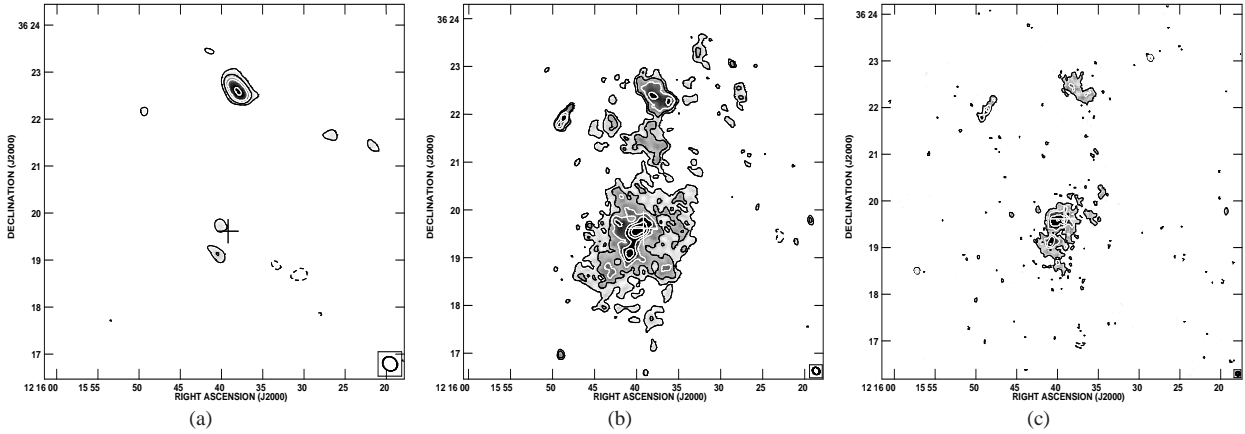

\subfigure[]{\includegraphics[height=5.5cm,width=5.5cm,angle=-90]{4214.150.ROB0.ps}\label{fig3:sub3}}
\subfigure[]{\includegraphics[height=5.5cm,width=5.5cm,angle=-90]{4214.330.ROB0.ps}\label{fig3:sub2}}
\subfigure[]{\includegraphics[height=5.5cm,width=5.5cm,angle=-90]{4214.610.ROB0.ps}\label{fig3:sub1}}
\caption{\small{The radio contour images of galaxy NGC 4214 with robust
weighting 0 are presented.  The radio contours are overlaid on radio grey
scale.  (a) 150 MHz, beam size = $21''.7\times17''.8$,  p.a.= $71^{\circ}.7$, the
contours are 3.5$\times$(-3, 3, 4.2,6,8.4,12) mJy/beam; 
(b) 325 MHz, beam size = $10''.4\times8''.6$, p.a.= $60^{\circ}.0$,
the contours are 0.34$\times$(-6, -3, 3, 4.2, 6, 8.4, 12) mJy/beam, ; 
(c) 610 MHz, beam size = $5''.4\times4''.8$, p.a.= $47^{\circ}.2$, the contours are 96 $\times$(-3, 3, 6, 12, 24) $\mu$Jy/beam.}}
\label{fig3:4214.ROB0}
\end{figure*}

The images of NGC 4214 at 150, 325 and 610 MHz made with robust weighting= 0
are shown in Figure~\ref{fig3:4214.ROB0}. 
The  325 MHz image, which has an rms noise of 0.34 mJy/beam, detects the
emission from both the central region and the diffuse halo  .
The 150 MHz detects only the compact emission from the central parts of the
galaxy while the 610 MHz image has picked up emission from diffuse regions
immediately surrounding the central compact regions. 
In Figure~\ref{fig3:sub2} the radio emission is observed to extend towards the
north from the main disk of the galaxy along a narrow ridge which connects to
the northern region.  The emission at 325 MHz appears to be fragmented at
several locations in the main disk of the galaxy, some of which could be
imaging artifacts.  We do not detect the diffuse extended emission at 150 MHz.
To understand this let us note that the 1 sigma rms noise in the image is 3.2
mJy/beam for a beam size of about $22''\times18''$, whereas the brightness   at
150 MHz, arrived at after employing the brightness of the diffuse region at 325
MHz and a spectral index of $-$1, is expected to  be 3.3 mJy beam$^{-1}$ and
which is below the 3 sigma sensitivity of our 150 MHz image.  As shown in
Figure~\ref{fig3:4214.ROB0}, we detect the intense star forming regions NGC 4214-I (northern) and
NGC 4214-II (southern) in the centre of NGC 4214.
We also analysed the 1.4 GHz VLA archival data (AM236) observed in A
configuration and have detected the star-forming regions NGC 4214-I and NGC
4214-II in addition to another star-forming complex in the north-west of these
regions.  Moreover at all the three bands, emission is detected from a region
located in the north around right ascension of $12^{h}15^{m}40^{s}$ and
declination of $36^{\circ}23'$.

NVSS images show extended emission from this galaxy.  We made a low resolution
map at 325 MHz to match the resolution of the NVSS.  These maps, were then used
to generate a spectral index distribution across the galaxy. The spectral index
across the galaxy varies from $\sim -1.5$ in the outer parts of the halo
emission to $\sim -0.5$ in the central parts of the galaxy.

NGC 4190 ($\alpha = 12^h13^m44.8^s$ , $\delta = +36^{\circ}38'03''$(J2000))
lying within the field of view of NGC 4214 is detected at 325 MHz with a flux
density of 8.1$\pm$1.0 mJy.

\subsection{NGC 4449}

\begin{figure*}
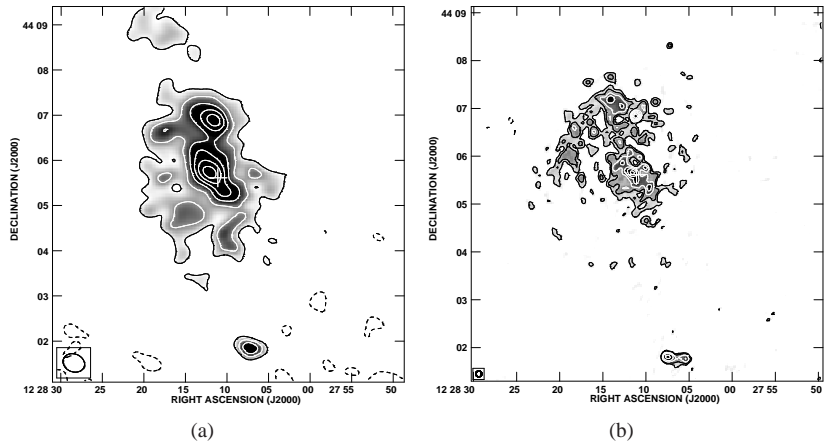

\subfigure[]{\includegraphics[height=5.5cm,width=5.5cm]{4449.150.ROB5.ps}\label{fig15.sub3}}
\subfigure[]{\includegraphics[height=5.5cm,width=5.5cm]{4449.610.ROB5.ps}\label{fig15.sub1}} 
\caption{The radio continuum maps of NGC 4449 made using robust
weighting of 5 with full uv range are presented here. The radio contours are
overlaid on its radio grey scale. (a) 150 MHz contours are 4.0 $\times$(-3,
3,5,7,9,11,13,15) mJy/beam, beam size = $28''.8\times23''.6$ , p.a. =
$65^{\circ}.4$  ; (b) 610 MHz contours are 0.21 $\times$(-6,6,8.4,12,16.9,
24,33.9,48) mJy/beam, beam size = $7''.6\times7''.1$, p.a. = $5^{\circ}.2$.}
\label{fig15:4449}
\end{figure*}

The 150 and 610 MHz images made with robust=5 weighting which are sensitive to
the extended emission are shown in Figure~\ref{fig15:4449}. A spiral arm-like
extension from east to west in the north of the galaxy is seen
(Figure~\ref{fig15.sub1}).  Diffuse radio emission is detected with two radio
peaks embedded within it.  The southern compact source is the centre of the
galaxy whereas the northern source is the supernova remnant J1228+441.  The
fragmented emission is likely due to imaging artifacts introduced by the sparse
longer baselines in the data.  
The presence of a strong source located about 7$'$ to the southwest of the
galaxy resulted in enhanced artifacts in our images.  

The double source seen to the south of NGC 4449 in both the 610 and 150 MHz
images is most likely a distant radio galaxy as suggested by \citet{chyzy2000}. 

\begin{table}
\centering
\caption{Flux Densities for NGC 4214 and NGC 4449}
\label{tab7:Flux Densities}
\begin{tabular}{@{}cccc@{}}
\hline
$\nu$ (MHz) &  NGC 4214               &  NGC 4449             & References          \\
            &  $S_{\nu}$ (mJy)        &  $S_{\nu}$ (mJy)      &                     \\
\hline                                                      
   150      & $104.1\pm15.6^{\#}$     & $639\pm95.8^{\#}$     & 1,1  \\
   325      & $192.5\pm43.9^{\#}$     & $785\pm225^{\#}$      & 1,1    \\
   609      &                         & $480\pm10$            & 2      \\
   610      &  $74.6\pm11.2^{\#}$     & $370\pm55.5^{\#}$     & 1,1    \\
   1400     &  $56.9\pm0.4$           & $272\pm0.4^{\#}$      & 3,3\\
   1400     &  $51.5\pm10.3^{\#}$     &                       & 4 \\
   1400     &  $38.3\pm7.7$           &                       & 5\\
   1400     &  $70\pm25$              &                       & 6\\
   1415     &                         & $224\pm22$            & 7\\
   1600     &  $65\pm25$              &                       & 6\\
   2380     &  $36.0\pm3.0^{\#}$      &                       & 8\\
   2700     &                         & $167^{\#}$            & 9\\
   3300     &  $\leq50$               &                       & 6\\ 
   4750     &                         & $138\pm10$            & 10\\
   4850     &  $30.0\pm4.5$           &                       & 11 \\
   4850     &  $30.0\pm7.0^{\#}$      &                       & 12\\ 
   4860     &  $34.0\pm6.8$           &                       & 4  \\
   5000     &                         & $135\pm16^{\#}$       & 13  \\ 
   8460     &  $20.5\pm0.5$           &                       & 14\\
   8460     &  $24.2\pm4.8^{\#}$      &                       & 4\\  
   10550    &                         & $87\pm6^{\#}$         & 2\\
   10770    &                         & $92\pm10$             & 15\\
   10770    &                         & $83\pm7$              & 10\\
   24500    &                         & $67\pm10^{\#}$        & 10\\
\hline \\
\end{tabular}
\\
\footnotesize{ The estimated systematic errors are quoted here.\\
{\#} shows the datapoints plotted in Figure~\ref{fignew10:4449}  \\
References:(1) This paper; (2) \citet{klein1996}; (3) NVSS; (4) \citet{kepley2011}; 
(5) \citet{condon2002}; (6) \citet{mas-Hesse1999} Radio data taken at Nancay Radio Telescope; 
(7) \citet{hummel1980}; (8) \citet{dressel1978}; (9) \citet{haynes1975}; 
(10) \citet{klein1986}; (11) \citet{becker1991}; (12) \citet{gregory1991}; 
(13) \citet{sramek1975}; (14) \citet{schmitt2006}; (15) \citet{israel1983}}
\end{table}

NVSS detects extended emission enclosing the entire disk of NGC 4449.  We used
the NVSS 1.4 GHz map of NGC 4449, which has an angular resolution of 45$''$ and
150 MHz (robust=5) image after smoothing the latter to the NVSS resolution to
generate a spectral index map of the radio emission.  The spectral index across
the galaxy varies from $-$1.5 in the outer regions to $-$0.5 in the inner
regions.

\begin{figure*}
\subfigure[]{\includegraphics[height=6.5cm,width=6.5cm]{4214GALEXFUV.PS}\label{fig6:4214GALEXFUV}}
\subfigure[]{\includegraphics[height=6.5cm,width=6.5cm]{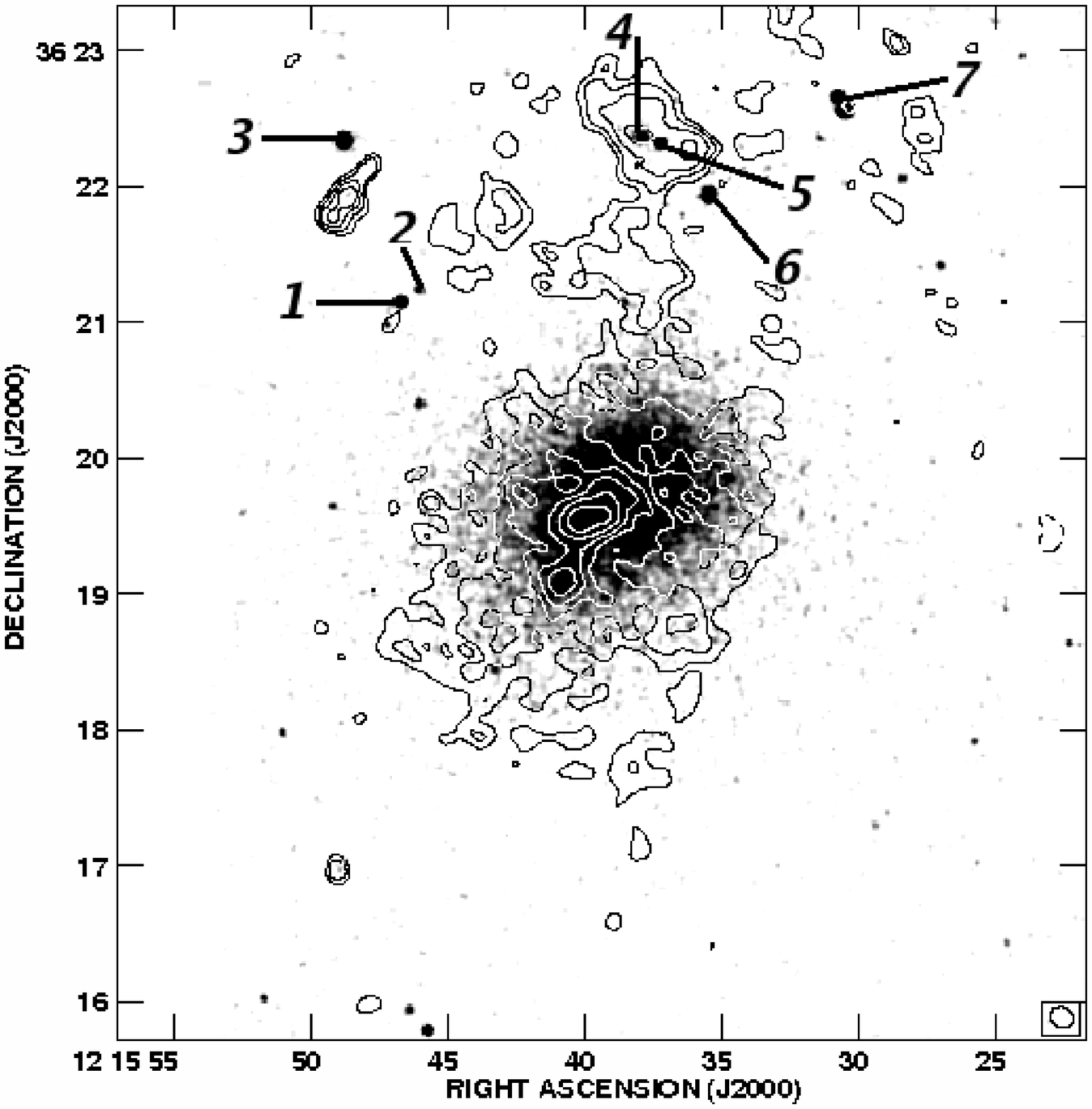}\label{fig7:4214.2MASS}}
\caption{NGC 4214, 325 MHz radio contour are superposed on (a)  GALEX-FUV gray
scale (b) 2MASS gray scale in which the labelled numbers represents the sources
in SDSS as: (1) star, (2) star cluster, (3) star, (4) 2MASX galaxy 2MASX
J1253795+362218 (5) NGC 4214:[HSS2004]X-rayS, (6) star, (7) star.  The contour
level is same as in Figure~\ref{fig3:sub2}.}
\end{figure*}

\subsection{Estimating the thermal emission}

The continuum emission at radio frequencies is mainly due to two emission
processes - namely synchrotron (also referred to as nonthermal) emission from
relativistic electrons accelerated in an ambient galactic magnetic field and
thermal free--free emission from star forming H{\sc ii} regions.  Owing to the
steep spectral index, of the nonthermal emission ($\ant< -0.5$) compared to the
thermal emission which has a spectral index of $-0.1$, it is expected to
dominate at low ($\lesssim1$ GHz) radio frequencies. Relativistic electrons are
believed to be accelerated in supernova shock fronts by diffusive shock
acceleration mechanism and are injected into the interstellar medium (ISM) with
$\ant\sim-0.5~{\rm to}-0.7$ \citep{bell1978, bogdan1983, biermann1993}. This
has been confirmed by observations of galactic supernova remnants
\citep[see][]{kothes2006, green2009}.  The electrons, then, propagate away and
lose energy in the process mainly due to synchrotron and/or inverse-Compton
cooling and thereby changing $\ant$ \citep[see e.g.][]{longair2011}. The study
of $\ant$ can throw meaningful insight into the energy loss/gain mechanisms of
relativistic electrons giving rise to the nonthermal emission. However, due to
the presence of thermal emission, the global spectral index $\alpha$ is
contaminated and is flatter than actual, particularly in giant H{\sc ii}
regions harbouring recent star formation activity.  

This, therefore, necessitates separating the thermal emission from the total
emission in these galaxies which have massive ongoing star formation
\citep{walter2001, reines2008}.  We estimate the thermal emission from these
galaxies, employing H$\alpha$ emission, following \citet{tabatabaei2007}, since
H$\alpha$ and thermal free--free emission arise in the same ionised gas. The
thermal emission flux density ($S_{\rm \nu, th}$) at a frequency $\nu$ is given
by, 
\begin{equation} 
S_{\rm\nu,th} = \frac{2k T_{\rm e} \nu^2}{c^2}(1 - e^{-\tau}) 
\end{equation} 
Here,
$k$ is the Boltzmann constant, $c$ is the velocity of light, $T_{\rm e}$ is the
temperature of thermal electrons assumed to be $10^4$ K and $\tau$ is the
free--free optical depth. $\tau$ is related to the emission measure ($EM$) as
\begin{equation} 
\tau = 0.082 T_{\rm e}^{-1.35} \left(\frac{\nu}{\rm GHz} \right)^{-2.1}
(1+0.08)\left(\frac{EM}{\rm cm^{-6}pc}\right)
\end{equation}
 The $EM$ is determined from the
H$\alpha$ intensity ($I_{\rm H\alpha}$) of the galaxy using equation (9) in
\citet{valls-gabaud1998}, \begin{equation} I_{\rm H\alpha} = 9.41\times10^{-8}
T_{\rm e4}^{-1.017}10^{-0.029/T_{\rm e4}} \frac{EM}{\rm cm^{-6} pc} ~~~{\rm
erg~cm^{-2}s^{-1}sr^{-1}} \end{equation} Here, $T_{\rm e4}$ is the electron
temperature in units of $10^4$ K. 

Thermal emission was computed for NGC 4214 and NGC 4449 for each pixel from the
H$\alpha$ maps observed with the 2.3-meter BOK telescope at Kitt Peak National
Observatory, downloaded from the NED. The galaxies were observed at a centre
wavelength of 6850 \AA, with a filter width of 68.7 \AA. The counts per second
in these maps were converted to apparent magnitude $m_{\rm AB}$ using the
zero-point given in the FITS file header. The specific intensity ($f_{\nu}$) was
calculated from $m_{\rm AB}$ i.e. $f_\nu ({\rm erg~s^{-1}cm^{-2}Hz^{-1}}) =
10^{-(m_{\rm AB}+48.6)/2.5}$, which was converted to the desired flux units of
$\rm erg~s^{-1}cm^{-2}$ from, $f = f_\nu d\nu = f_\nu c (d\lambda/\lambda)^2$.
The thermal emission map determined based on the H$\alpha$ maps has an
angular resolution of $\sim2''$. The optically thin free--free emission was then 
removed from the total flux density at all frequencies, and the remnant flux was 
assumed to be due to synchrotron emission. Integrated radio spectrum of NGC 4214 and 
NGC 4449 are  displayed in Figure~\ref{4214integ.ps} and \ref{4449integ.ps} where  solid, dot-dashed and dashed lines 
represent, respectively, the total, the nonthermal, and the the thermal components. 
A power law fitting gives us $\alpha_{\rm nt} = -0.63\pm0.04$ for NGC 4214 and $\alpha_{\rm nt} = -0.49\pm0.02$ for NGC 4449.  

We note that the H$\alpha$ emission suffer due to extinction from dust present
in these galaxies. The dust extinction was estimated, using the optical depth,
$\tau_{\rm dust}$, of the obscuring dust at FIR wavelength of $\lambda160~\mu$m
\citep{tabatabaei2007}.  We used the MIPS data of the Spitzer space telescope
to estimate the dust temperature and thus $\tau_{\rm dust}$. $T_{\rm dust}$ was
estimated by fitting a modified black body spectrum of the form, $f_{\rm IR}
\propto \nu^\beta B_{\rm IR}(T_{\rm dust})$ between $\lambda$70 and 160 $\mu$m,
where $B_{\rm IR}(T_{\rm dust})$ is the Planck function.  Here, $f_{\rm IR}$ is
the flux at a FIR frequency $\nu_{\rm IR}$ and $\beta$ is the power-law index
of dust absorption efficiency assumed to be 2 \citep{draine84}. The optical
depth at H$\alpha$ wavelength was estimated by extrapolating $\tau_{\rm dust}$
using the standard dust model for our galaxy \citep{kruegel2003}.  In dense
H{\sc ii} regions, the extinction was found to be $\sim20-30\%$, while in other
parts it was $<10\%$.  Thus, the H${\alpha}$ flux and thereby the thermal
emission at radio frequencies determined by us in this study are lower limits.

\section{Discussion}

\subsection{NGC 4214}

\subsubsection{Morphology and correlation with emission at other wavebands}

\begin{figure*}
\subfigure[]{\includegraphics[height=7cm,width=7cm,angle=-90]{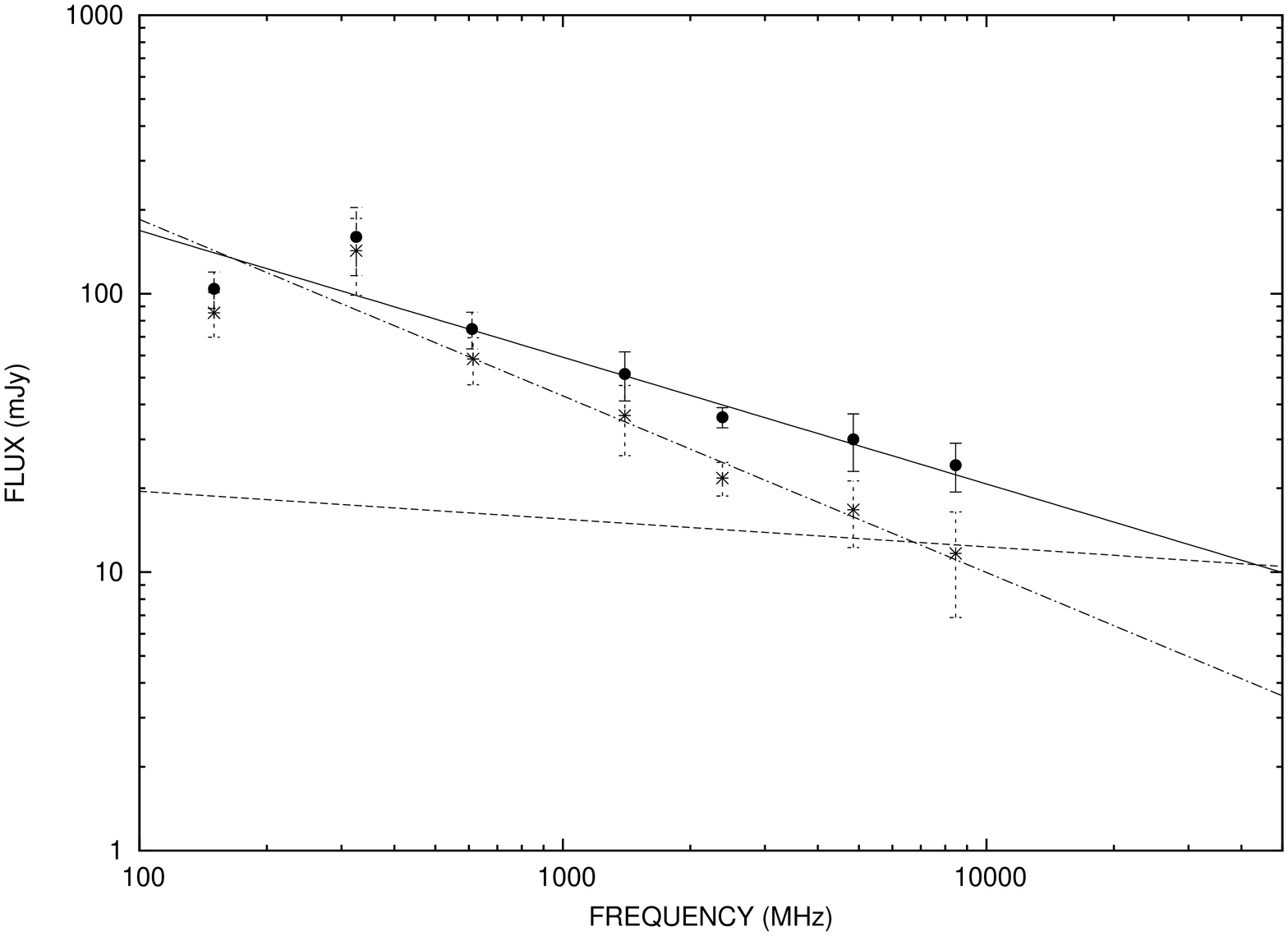}\label{4214integ.ps}}
\subfigure[]{\includegraphics[height=7cm,width=7cm,angle=-90]{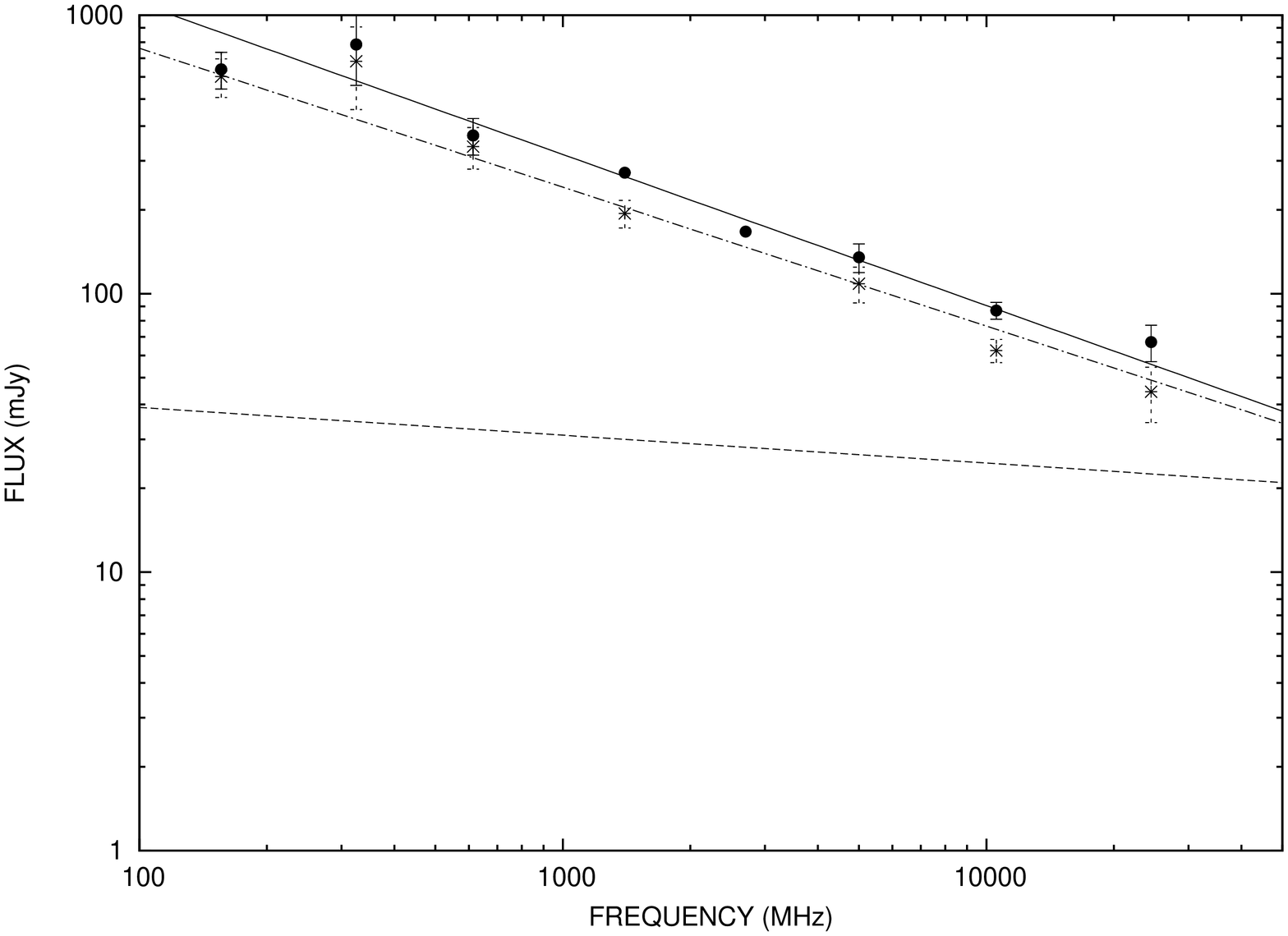}\label{4449integ.ps}}
\caption{Integrated radio spectrum of (a) NGC 4214 and (b) NGC 4449. In both
the spectra the data points $<$1.4 GHz are from this paper.  The solid line
represents the total, the dot-dashed the nonthermal, and the dashed line the
thermal components.  The fitting for the total and non-thermal was done
excluding the points at 325 and 150 MHz for NGC 4214 whereas for NGC 4449 data
point at 325 MHz was excluded.  Thus by a power law fitting we get $\ant$ = $-0.63\pm0.04$ for NGC
4214 and $\ant$ = $-0.49\pm0.02$ for NGC 4449. The details of the plotted data
are given in Table \ref{tab7:Flux Densities}. }
\label{fignew10:4449}
\end{figure*}

In Figures~\ref{fig6:4214GALEXFUV} and ~\ref{fig7:4214.2MASS}, the radio
contours at 325 MHz are shown superposed on the GALEX\footnote{Galaxy Evolution
Explorer is an orbiting ultraviolet space telescope.} FUV (1350 - 1750\AA) and
2MASS NIR images.  Scattered star--forming regions are seen to the west, south
and north of the intense star forming disk of NGC 4214 in the UV images,
whereas the NIR emission has a boxy morphology.  Radio emission at 325 MHz is
detected from the entire FUV disk.  The angular extent of the 325 MHz halo is
about 2.3 kpc.  While radio emission is detected from a part of the northern
star forming region, no radio emission is seen to be associated with the
ultraviolet emission in the west and south of the disk of the galaxy.  
We note that there are two more discrete radio continuum regions in the field,
similar to as pointed out earlier by \citet{kepley2011} - one in the northern
part of NGC 4214 and the other in the northeast.  While FUV emission is
associated with the western part of the northern star forming region, no such
emission is associated with the source in the north-east.  Our high-resolution
images (e.g. Figure~\ref{fig3:sub1}) show the source in the north-east to be a
double source and our results support the conclusion by \citet{kepley2011}
about it being a background radio galaxy.  

The NUV and FUV emissions show the galaxy to be undergoing vigorous star
formation, which extends along a bridge-like feature to the north.  Radio
emission at 325 MHz is also detected along the bridge-like extension and the
northern star forming region.  The SFR estimated from the non-thermal emission
is $1.97\times10^{-8}~{\rm M_{\odot} yr^{-1}pc^{-2}}$.  The UV emission appears
to be of irregular morphology, with the star formation triggered at various
locations along a thick and almost north-south ridge.  While the extent along
the major axis of the galaxy is similar, the radio continuum appears to be boxy
in nature (Figure~\ref{fig7:4214.2MASS}).  
The old stellar population is distributed in a spherical halo resembling a
large globular cluster as is also seen in the I band image of
\citet{fanelli1997}.  The radio continuum appears to be more extended than the
NIR emission.  No NIR emission is detected along the bridge-like extension or
from the northern star-forming region, indicating a fairly recent star
formation episode in these regions.  The radio emission associated with this
northern region is more extended in the east compared to the FUV (see
Figure~\ref{fig6:4214GALEXFUV}).  Two radio peaks are visible in this northern
region at 610 and 325 MHz, with the eastern peak coinciding with the 2MASS
source, 2MASX J12153795$+$3622218, which is classified as a background galaxy
(see Figure~\ref{fig7:4214.2MASS}).  The western radio peak region shows the
presence of FUV emission and is a star-forming region in NGC 4214. At 150 MHz
the emission from the northern region is seen to be exceptionally bright in
comparison with the rest of the galaxy. The spectrum of this northern region
between 150 and 610 MHz is fitted by a single power-law with a spectral index
of $-1.1$.  X-ray emission has been detected from this region by
\citet{ott2005}. The emission at 150 MHz shows a single peak coincident with
the 2MASS galaxy.  
Thus, we suggest that the northern region consists of two distinct parts - the
western part which is associated with the star-forming region in NGC 4214 and
the eastern part which is associated with the background galaxy 2MASX
J1253795+362218 \citep*{srivastava2011}.  The two peaks are easily identifiable
in our 610 MHz image in Figure~\ref{fig3:sub1}.  The radio galaxy to the
north-east and the galaxy in the north are likely part of a larger distant
cluster.  

An alternative scenario is provided by \citet{kepley2011} who suggest that the
entire northern part is a star-forming region of age $10-13.5$ Myr and the
emission at 20 cm is elongated due to a uniform magnetic field with a strength
of 7.6 $\mu$G.
The central star forming regions are well resolved and compare well with the
higher radio frequency maps by \citet{beck2000}.  The molecular gas has been
reported confined to three distinct regions - NGC 4214-I, NGC 4214-II and and a
region $\sim$ 760 pc to the north-west \citep{walter2001}. Star-forming regions
are close to all the three clouds and radio continuum emission is detected from
these.  Starbursting dwarf galaxies tend to be H{\sc i} - rich
(\citealt{thuan1981}; \citealt{taylor1995}).  The H{\sc i} from NGC 4214 is
reported to be extended \citep{allsopp1979}.  However, the radio continuum at
325 MHz is found to bear little resemblance to this.

\subsubsection {The global spectrum}

In this subsection, we discuss the global spectrum of NGC 4214 after combining
our three-frequency radio data with data from literature.  The data are  listed
in Table~\ref{tab7:Flux Densities} and the entries marked by a \# are plotted
in Figure~\ref{4214integ.ps}.  The integrated spectral index down to 610 MHz is
fitted by a power-law spectrum of index $-0.45\pm0.03$.  This is similar to the
value of $- 0.43\pm0.06$ determined by \citet{kepley2011} for $\nu >$1 GHz.
The presence of the halo emission detected at 325 MHz makes the galaxy similar
to normal star-forming disk galaxies and NGC 4449 \citep{klein1996}.  We note
that the low frequency spectrum includes a contribution from the region located
to the north of the galaxy.  

Integrated radio spectrum presented in Figure~\ref{4214integ.ps} reveals that the thermal
emission starts becoming significant at higher radio frequencies,  with about
{40\%} of the total emission near 5 GHz and 22{\%} at 610 MHz being thermal in
origin.  This is significantly higher than $\sim$10$\%$ thermal emission at 1
GHz found for normal star-forming galaxies.  We determine the synchrotron flux
density by subtracting the thermal emission from the total flux density at all
frequencies of observation and represent it in Figure~\ref{4214integ.ps} by
dot-dash line.
It is pointed out that the flux density is high at 325 MHz due to contribution
from diffuse halo emission and is low at 150 MHz due to sensitivity issues,
hence, we excluded these points from the power-law fit to the non-thermal
spectrum and
and we estimated $\alpha_{\rm nt}$ to be $-0.63\pm0.04$ for NGC 4214.  Thus a
combination of synchrotron spectrum with index $-0.63$ and free-free thermal
emission with index $-0.1$ explains well the observed spectrum between 610 MHz
and 8.7 GHz (Figure~\ref{4214integ.ps}).
From the spectral index map made between our 325 MHz and 1.4 GHz (NVSS), we
find that the spectrum of the halo emission in the outer parts of the galaxy is
$\sim -1.5$, indicative of an aged electron population.    

\subsubsection{The compact star forming regions: NGC 4214-I and NGC 4214-II}

As shown in Figure~\ref{fig3:4214.ROB0}, we detect the intense star forming
regions in the centre of NGC 4214: NGC 4214-I and NGC 4214-II. These regions
are detected at 325 MHz as surrounded by diffuse extended.  To study these
compact regions, we imaged the GMRT data by excluding baselines shorter than
3$k\lambda$, so that the extended emission was resolved (see
Figure~\ref{4214HIGRES}). 
We study the spectra of these two regions from 325 MHz to 15 GHz after
including the high frequency data between about 4 to 15 GHz from
\citet{beck2000} (see Figures~\ref{4214HIGRES}, \ref{radio.spec}). 
As noted in the previous section, the low sensitivity of the 150 MHz image did
not allow detection of the diffuse emissions from these regions.  Besides, the
flux densities of NGC 4214-I and NGC 4214-II are also lower than predicted by a
non-thermal spectrum (see Figure~\ref{radio.spec}), implying that thermal
absorption could also be a reason for non-detection of the diffuse emissions.
Since it is difficult to disentangle the two effects with the current data, we
have excluded the 150 MHz point from our analysis. The spectra of both the
regions display similar behaviour but significantly different from the global
spectrum (see Figure~\ref{4214integ.ps}), especially at high radio frequencies.
The spectra shown in Figure~\ref{radio.spec} appear to consist of three
components: the synchrotron non-thermal emission dominating at the lowest
frequencies, the optically thin $(\tau < 1$) free-free thermal emission and the
optically thick ($\tau >1$) free-free thermal emission dominating at the higher
frequencies.
We note that the last component ($\tau >1$) has been studied by
\citet{beck2000} who suggested a turnover frequency for this component to be
$\sim 15$ GHz.  We have tried to ensure that all the data points used in our
analysis (see Figure~\ref{radio.spec}) were sensitive to the largest angular
scale that we are probing for the two regions.  We have excluded 8 GHz data
point \citep{beck2000} from our analysis on the ground that the largest angular
scale probed by them was 15$''$ and hence, it is likely that the flux is missed
out.

In the first step of the analysis, the optically thin thermal emission was
estimated by extrapolating from the H$\alpha$ map of the region.  No correction
for internal extinction was made to the H$\alpha$ map.  The estimated spectrum
is shown by the dashed lines in Figure~\ref{radio.spec}.  We, then subtracted
this thermal emission from the total flux density at each frequency.  In the
second step, the remaining emission was fitted by a combination spectrum of the
type:

\begin{figure*}
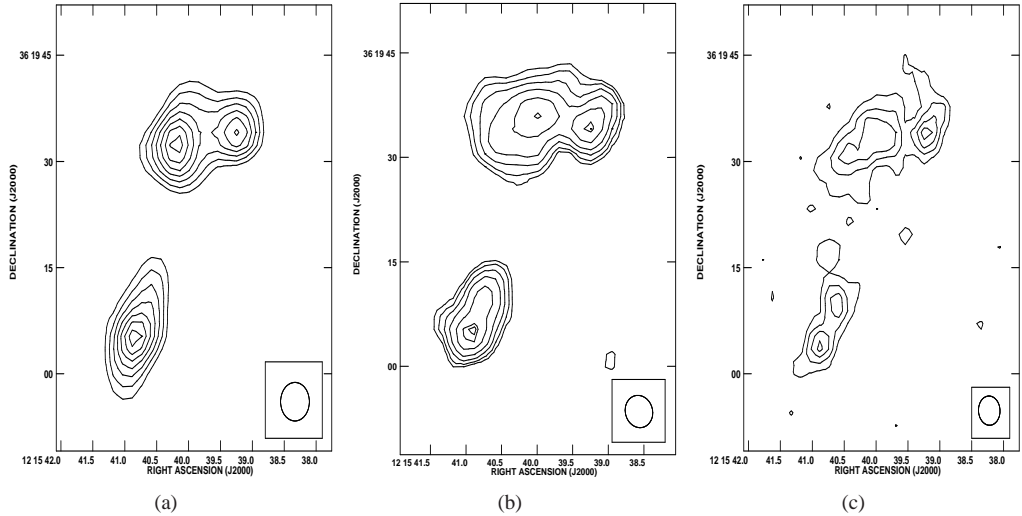

\begin{center}
\subfigure[]{\includegraphics[height=6.5cm, width=4.5cm]{4214.330.LOWRES.ps}\label{fig11.sub2}}
\subfigure[]{\includegraphics[height=6.5cm, width=4.5cm]{4214.610.LOWRES.ps}\label{fig11.sub1}}
\subfigure[]{\includegraphics[height=6.5cm, width=4.5cm]{4214.1400.LOWRES.ps}\label{fig11.sub3}}
\caption{Images showing NGC 4214-I and NGC 4214-II at (a) 325 MHz, contours are 0.3$\times$(3,4,5,6,7,8,9) mJy/beam; 
(b) 610 MHz; contours are 0.1$\times$(3,4.2,6,8.4,12,16) mJy/beam and (c) 1.4 GHz image from VLA
archival data, contours are 0.16 $\times$(-3, 3, 4.2, 6, 8.4) mJy/beam.  All
images are made by excluding baselines shorter than 3k$\lambda$.  }
\label{4214HIGRES}
\end{center}
\end{figure*}

\begin{figure*}
\begin{center}
\subfigure[]{\includegraphics[height=7cm,width=7cm,angle=-90]{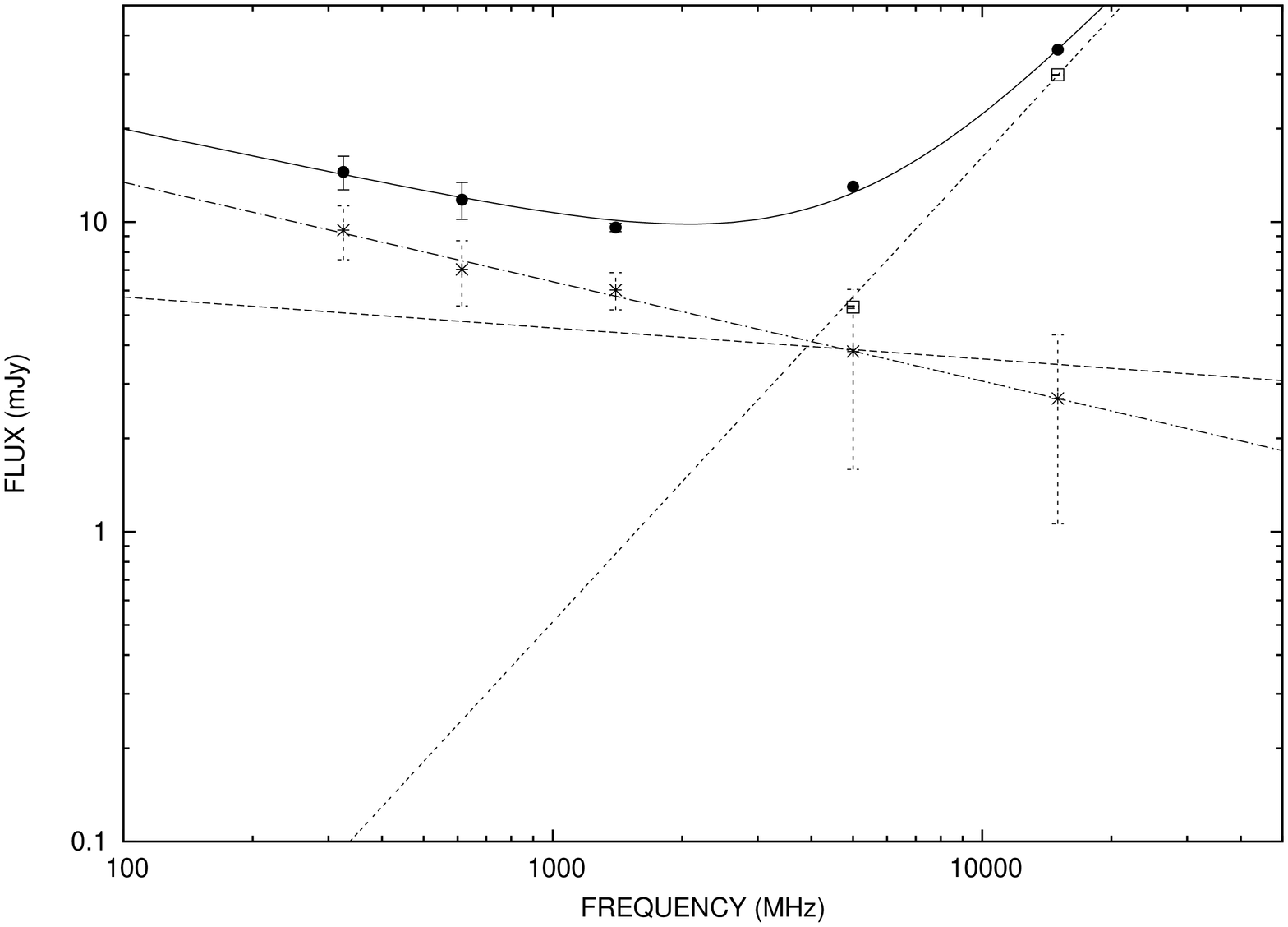}\label{radio.spec_sub2}}
\subfigure[]{\includegraphics[height=7cm,width=7cm,angle=-90]{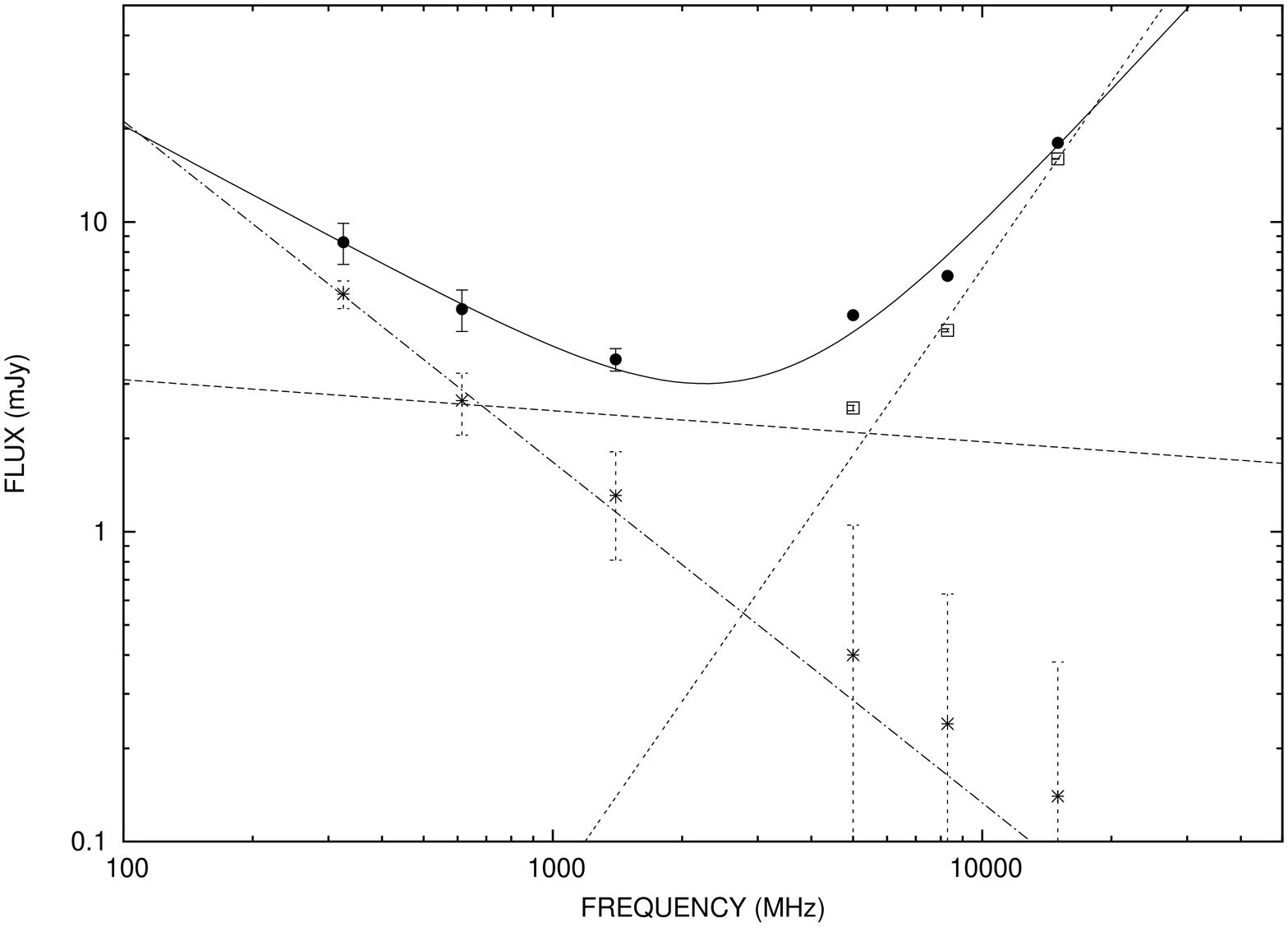}\label{radio.spec_sub3}}
\caption{The spectra of star-forming regions (a) NGC 4214-I;
$\alpha_{nth}=-0.32\pm0.02$ (b) NGC 4214-II;  $\alpha_{nth}=- 0.94\pm0.12$.
In both plots the solid line represents the final model values; the dashed
line shows the thermal spectrum estimated using  $S\propto\nu^{-0.1}$; the
dot-dash line represents the non-thermal emission fitted by the best fitting
power law and the filled square data points are estimated by subtracting the
thermal and non-thermal contribution from the total value and shows the
variation of the flux density as  $\nu^{2}$ relation (small dashed line). The
data points at frequencies less than 1 GHz are from the GMRT, the 1.4 GHz data
is from VLA archival data and the data points at frequencies $>$1.4 GHz are
from \citet{beck2000}.}
\label{radio.spec}
\end{center}
\end{figure*}

\begin{equation}
f(\nu)=a\nu^{\alpha_{nt}}+c\nu^{2}
\end{equation}
This simple model resulted in a reasonably good fit (solid line curve in
Figure~\ref{radio.spec}) to the data for both the complexes.
While the optically thick component was fixed to vary as $\nu^2$, the
non-thermal spectral index was a free parameter in the model.  The best fits
gave $\alpha_{nth}$ =$- 0.32\pm0.02$ for NGC 4214-I and $\alpha_{nth}$ =$-
0.93\pm0.12$ for NGC 4214-II.  

The thermal contribution due to the optically thin component for NGC 4214-I at
325 MHz, 610 MHz and 1.4 GHz are about 35\%, 40\% and 42\% whereas  for NGC
4214-II, it is about 32\%, 49\% and 54\% respectively.   \citet{kepley2011}
also estimate the thermal emission at 1.4 GHz in central parts to be about 50\%
of the total emission.  The complex spectrum of NGC 4214-I and II indicate the
presence of $\tau >$1 ultra-compact HII regions, $\tau<$1 diffuse thermal and
non-thermal emission.

\begin{figure}
{\includegraphics[height=8.5cm,width=10cm]{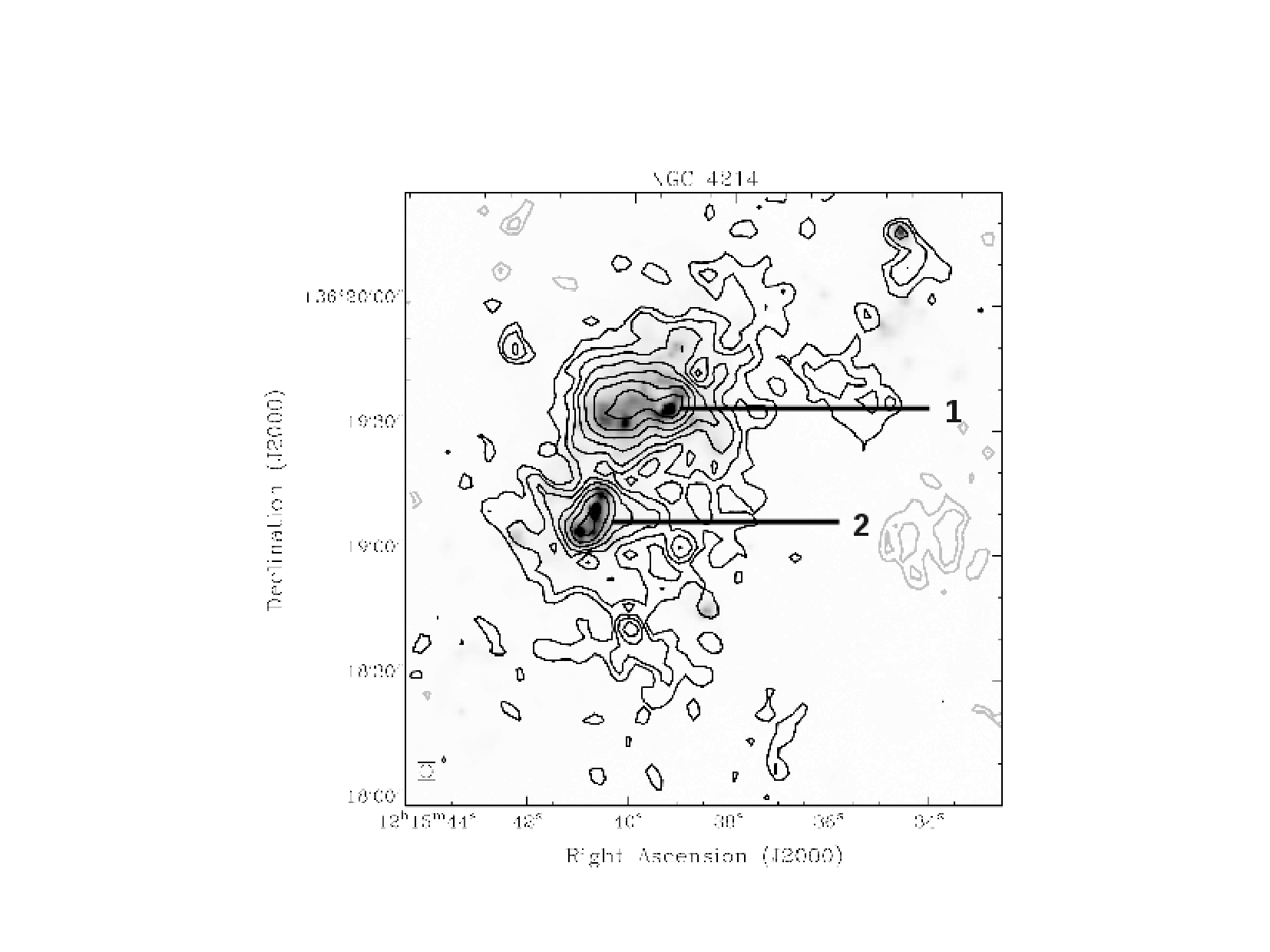}}
\caption{Radio contours showing the 610 MHz emission
overlaid on the H$\alpha$ image of NGC 4214.  
$1$ shows NGC 4214-I and $2$ shows NGC 4214-II}
\label{ngc4214Halpha}
\end{figure}

A detailed look at Figure~\ref{4214HIGRES} shows two radio peaks in NGC 4214-I,
the eastern peak is diffuse and intense at 325 MHz whereas the western peak is
strongest at 1.4 GHz.  This indicates a steeper spectrum in the east and hence
possibly an older star-forming region.  The H$\alpha$ image (see
Figure~\ref{ngc4214Halpha}) shows bright compact regions in both the eastern
and western half of NGC 4214-I surrounded by diffuse emission.  The eastern
H$\alpha$ peaks are displaced to the south of the radio peak at 610 MHz whereas
the western H$\alpha$ peak coincides with the radio peak.
We recall that such displacement is commonly seen in the circumnuclear rings in
disk galaxies and is believed to be due to the presence of dust.  Good
correlation between the location of the 610 MHz emission and the H$\alpha$
emission is also observed in NGC 4214-II (see Figure~\ref{ngc4214Halpha})
indicating lower dust extinction in this star forming region.  We note that
this galaxy is overall a dust-deficient galaxy \citep{mackenty2000} with the
stellar clusters in NGC 4214-I of age $\sim$3~-~4 Myr, whereas the clusters in
NGC 4214-II are younger at about 2.5~-~3 Myr.

The radio luminosity at 610 MHz is $1.2\times10^{19}$ WHz$^{-1}$ for NGC 4214-I
and $5\times10^{18}$ WHz$^{-1}$ for NGC 4214-II for a resolution of about 200
pc. The luminosities of these star-forming regions (10$^{19}$ W~Hz$^{-1}$)
appear to be similar to those in circumnuclear rings of typical diameter
$500-1000$ pc in several normal disk galaxies observed with similar linear
resolution.
For example, \citet{kodilkar2011}, who have done a radio study of the
circumnuclear ring in NGC 2997 with a linear resolution of 200 pc, find that
the luminosity of the five star-forming clumps resolved in the ring at 1.4 GHz
is between $10^{19}$ to $10^{20}$ W~Hz$^{-1}$.  Moreover, \citet{kodilkar2011}
estimate an equipartition magnetic field of about 30 $\mu$G for the
circumnuclear ring in NGC 2997 which is similar to that estimated for the
central regions of NGC 4214 by \citet{kepley2011}. Thus star-forming regions
in the centre of NGC 4214 appear to display properties similar to the centres
of normal disk galaxies.   This is important since NGC 4214 is a dwarf WR
galaxy of lower metallicity and it is instructive to compare their star-forming
complexes with those in the better studied disk galaxies. 

Finally, we comment on the possible corrections that can be included to improve
this study.  Firstly, the correction for extinction in the H$\alpha$ signal due
to dust in NGC 4214 can be included.
We estimate a maximum variation of 20\% in the estimates of thermal emission
due to extinction.
Including dust extinction might lead to a synchrotron spectrum that is steeper
than the estimated $-0.32\pm0.02$ for NGC 4214-I.  The effect of dust does not
seem to be observable in the H$\alpha$ image of NGC 4214-II (see
Figure~\ref{ngc4214Halpha}).
This would indicate that only small corrections  will have to be made on the
estimated non-thermal spectrum.  Secondly, the radio maps used to obtain this
interesting result are made with varying uv coverage of the fixed antenna
configuration of GMRT.  Being low radio frequencies, the data are subject to
extensive flagging.  Thus, there is a possibility that the images might be
sensitive to somewhat different angular extents.  

\subsection{NGC 4449}

This galaxy has been extensively studied in various wavebands including low
radio frequencies.  \citet{klein1996} studied the galaxy at 610 MHz with the
WSRT and detected an extended synchrotron halo around the optical galaxy.
\citet{chyzy2000} have detected the radio halo at 8.46 and 4.86 GHz and mapped
the galaxy in polarised emission. Our continuum observations trace the regions
of large scale regular magnetic field seen by \citet{chyzy2000}.  Extended
radio emission was detected at all the observed GMRT bands of 150 MHz, 325 MHz
and 610 MHz.  

\subsubsection{Comparing with emission at other wavebands}

\begin{figure*}
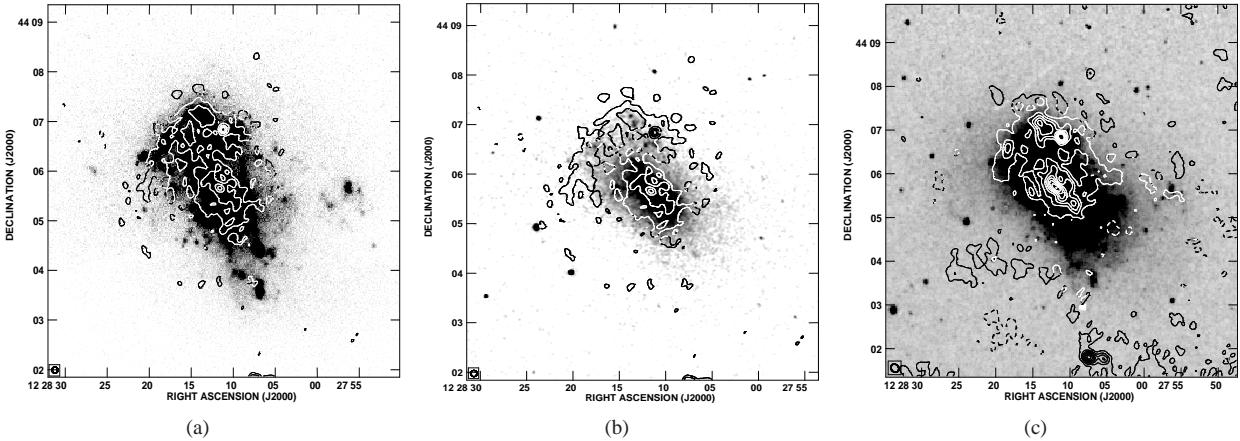

\subfigure[]{\includegraphics[height=5.5cm,width=5.5cm]{4449.610GALEXFUV.ps}\label{fig25:4449.610GALEXFUV}}
\subfigure[]{\includegraphics[height=5.5cm,width=5.5cm]{4449.6102MASS.PS}\label{4449.6102MASS}}
\subfigure[]{\includegraphics[height=5.5cm,width=5.5cm]{610_0TO20_ROB5_4449.ps}\label{0TO20_ROB5_4449}}
\caption{NGC 4449, 610 MHz, Rob =5 radio contour is superposed on (a) GALEX-FUV
gray scale, (b) 2MASS gray scale. The contour level is
0.21$\times$(-6,6,12,24,33,48) mJy/beam (c) 610 MHz
made with ROB 5 and uvcutoff of 20 k$\lambda$ superposed on the DSS grey scale
image.  Beam Size = $12''.30\times10''.40$, p.a. =$50^{\circ}.35$.
The contours are 0.41 $\times$(-3, 3, 6, 9, 12, 15, 18, 21) mJy/beam. }
\end{figure*}

The radio emission detected at 150 MHz and 610 MHz is shown in
Figure~\ref{fig15:4449}.  We estimate the size of the 150 MHz halo to be about
4 kpc which is similar to the size of 3.5 kpc that \citet{chyzy2000} estimate
but smaller than the 7 kpc that \citet{klein1996} estimate.  The spiral
arm-like feature in the north of the galaxy, clearly detected in our 610 MHz
images shown in Figures~\ref{fig25:4449.610GALEXFUV} and \ref{4449.6102MASS}
has highly ordered magnetic field \citep{chyzy2000, klein1996}. Polarization
study by \citet{chyzy2000} reveals intriguing galaxy-scale magnetic fields of
$6-8~\mu$G, even though the galaxy is a slow, chaotic rotator.

NGC 4449 is estimated to be about five times more luminous than NGC 4214,
indicating vigorous star formation in NGC 4449.  The star formation rates
estimated from the non-thermal emission is  $3.35\times10^{-8}~{\rm
M_{\odot}yr^{-1} pc^{-2}}$.  About 10 ULXs (X-ray luminosity $> 10^{39}{\rm erg
s^{-1}}$) have been detected in NGC 4449 \citep{ott2005} which confirms the
heightened level of activity in this galaxy.  We have overlaid the radio
contours at 610 MHz on the GALEX FUV map  in
Figure~\ref{fig25:4449.610GALEXFUV} and NIR image from 2MASS in
Figure~\ref{4449.6102MASS}.  The FUV map show an almost north-south
distribution of vigorous star formation, whereas the 2MASS emission is boxy in
nature and centrally concentrated.  This is similar to NGC 4214 discussed in
the previous section.  While the 2MASS emission shown in
Figure~\ref{4449.6102MASS} is symmetric about the centre of the galaxy, the
ultraviolet and radio emission are asymmetric, with excess emission detected in
the north of the galaxy.   The ultraviolet emission extends further south
compared to the radio continuum emission.  The spiral arm-like feature seen in
the north of the galaxy shows a counterpart in the ultraviolet.  However, it is
seen to skirt the NIR emission in the galaxy, indicating the recent 
star formation there.  The extent of disk in south is similiar to radio.  The
X-ray emission is surrounded by an extended HI envelope \citep[see
e.g.][]{yun1994, hunter1998}.

\subsubsection {The global spectrum }

Our multi-frequency data alongwith the integrated flux density data available
in literature are listed in Table~\ref{tab7:Flux Densities}.  
Figure~\ref{4449integ.ps} shows the global spectrum of NGC 4449 in the
frequency range of 150 MHz to 22 GHz.  \citet{chyzy2000} have presented
spectral-index maps between 8.46 GHz and 4.86 GHz and find that the spectral
index changes from  $\sim -0.3$ in strongly star forming regions to $\sim -1.1$
locally in the outer southern region.   They also reported that the
radio-bright peaks coincide with strongly star-forming regions which show
increased thermal fractions (up to 80$\%$ at 8.46 GHz).  \citet{klein1996},
estimated a thermal fraction at 1 GHz of $f_{th} = 10\pm4\%$ and a non-thermal
spectral index $\alpha_{nth}$ = $-$0.7$\pm$0.1.

Figure~\ref{4449integ.ps} reveals that about 20\% of the total emission near 5 GHz is thermal
in origin. The synchrotron spectral index for NGC 4449 using 150 MHz and 22 GHz
data is estimated to be $-0.49\pm0.02$.  The 325 MHz point was excluded.
Interestingly, \citet{klein1996} record a flux density of 480 mJy at 610 MHz.
The spectral index between this and our 325 MHz data is $-$0.74.  Thus a
combination of a synchrotron spectrum with index $-$0.49 and free-free thermal
emission with index $-$0.1 explains the observed spectrum between 150 MHz and
22 GHz as shown in Figure~\ref{4449integ.ps}.  We estimate that about 9\% of
the total emission at 610 MHz is thermal in origin, which is similar to normal
disk galaxies but less than to what we estimate for NGC 4214 ($22\%$).  One
possible reason could be the higher dust extinction in NGC 4449 leading to an
under estimation of the thermal fraction at radio frequencies from the
H$\alpha$ map.  Alternatively, the continuous massive star formation in NGC
4449 might be arising enhanced non-thermal emission in the galaxy.  The
spectral index is flatter than found in normal galaxies, suggesting continuous
star formation injecting energy into the relativistic plasma.

\subsubsection{The supernova remnant SNR J1228+441:} 

The flux density of the supernova remnant SNR J1228+441 was estimated from
images made after excluding the short baselines.  The SNR was unresolved in all
our images and is the most intense feature in NGC 4449 at all GMRT frequencies.
This SNR is five time more luminous than Cas A at 20cm \citep{chomiuk2009}.  It
has been extensively monitored (e.g.  \citet{lacey2007} and references
therein). \citet{lacey2007} report steepening of the spectral index from
$\alpha=-0.64\pm0.02$ in 1994 to $\alpha=-1.01\pm0.02$ in 2001-2002, showing
rapid evolution.  \citet{reines2008} using high--resolution VLA data at several
frequencies find that the supernova remnant had a spectral index of $-1.8$
between 3.6 cm and 6 cm  and an index of $-0.9$ between 1.3 cm and 3.6 cm
between 2001 and 2002, indicating a break in the spectrum.  From our
observations at 325 and 610 MHz in 2008-2009, we estimate flux densities of
35.2$\pm$9.1 mJy and 11.2$\pm$2.9 mJy, respectively, and at 150 MHz in 2012, we
estimate a flux density of 32.4$\pm$5.1 mJy.  The spectral index between 325
and 610 MHz is calculated to be $-1.8$ for epoch 2008-2009.   

\subsection{Radio--FIR correlation}

\begin{figure*}
\subfigure[]{\includegraphics[height=7cm,width=7cm]{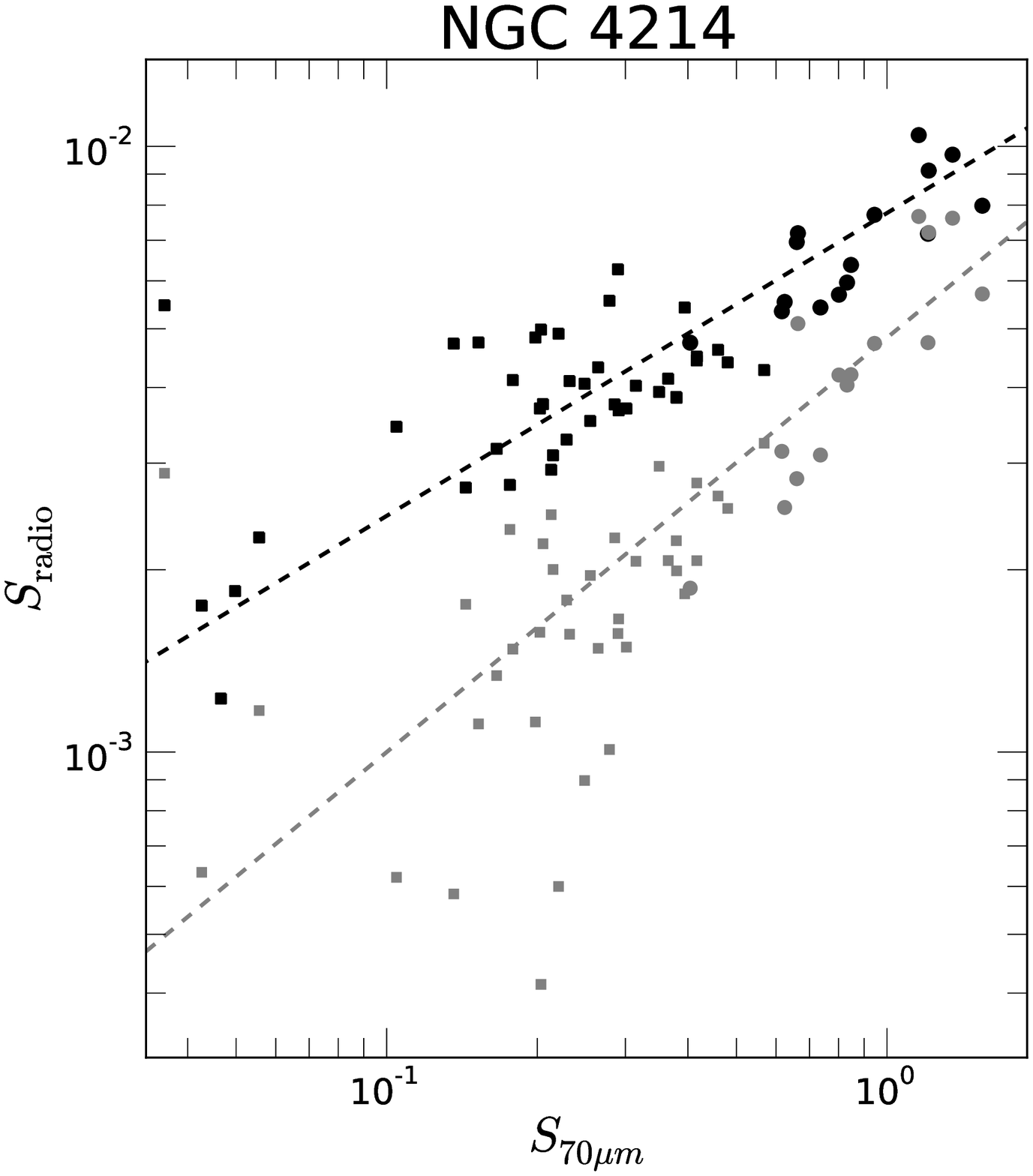}\label{rf1.4214}}
\subfigure[]{\includegraphics[height=7cm,width=7cm]{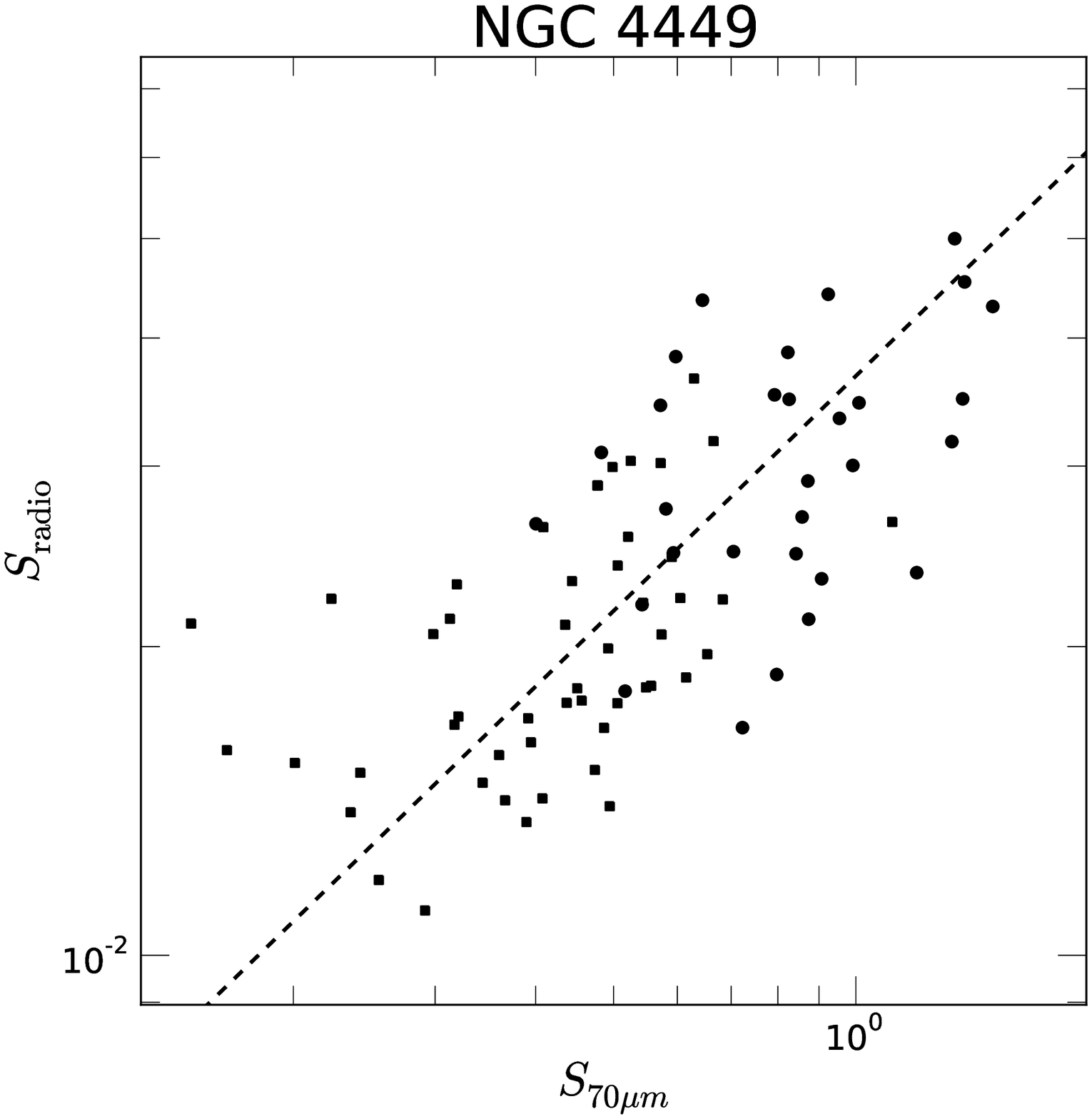}\label{rf1.4449}}
\subfigure[]{\includegraphics[height=5cm,width=5cm]{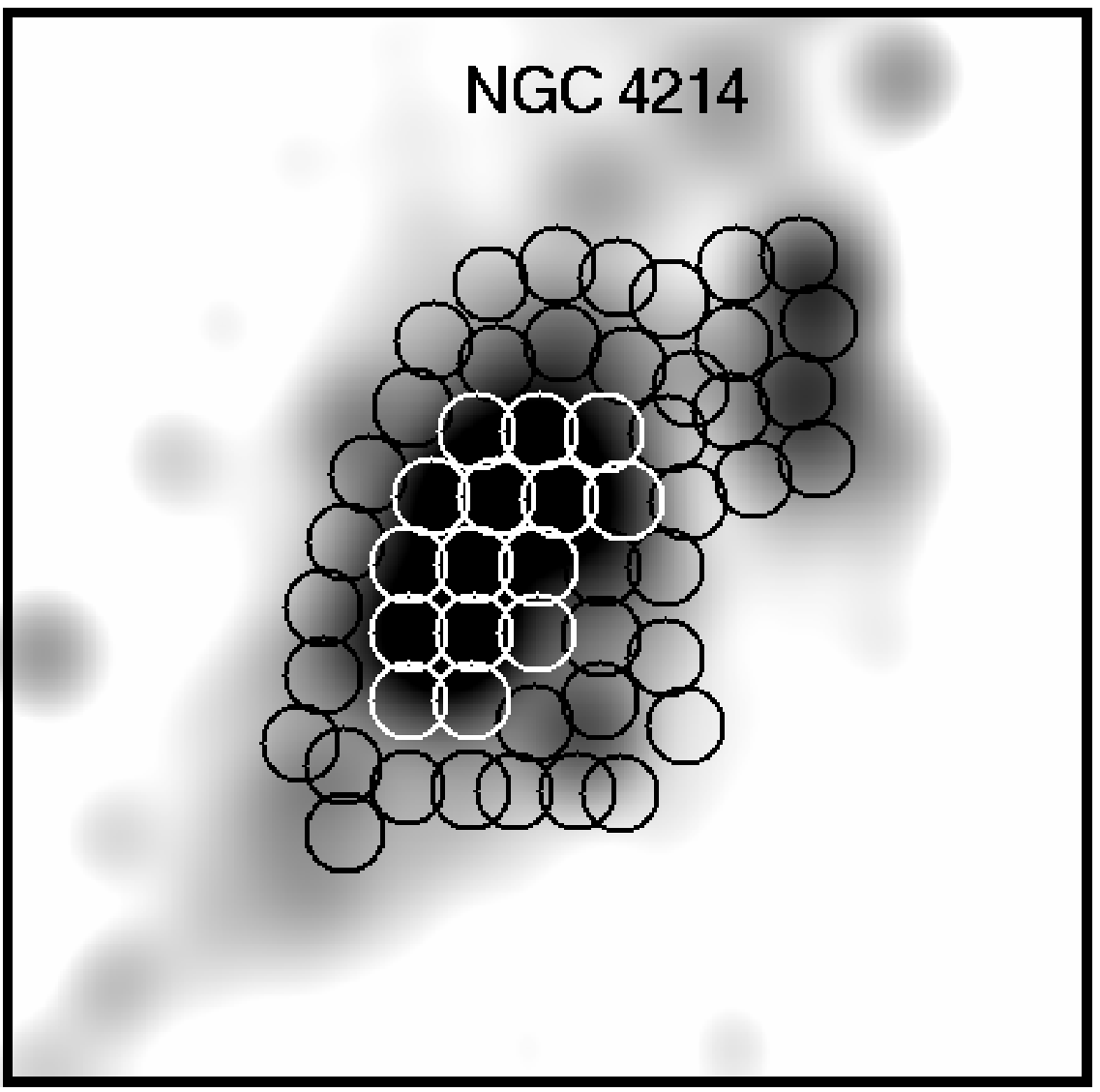}\label{rf2.4214}}
\subfigure[]{\includegraphics[height=5cm,width=5cm]{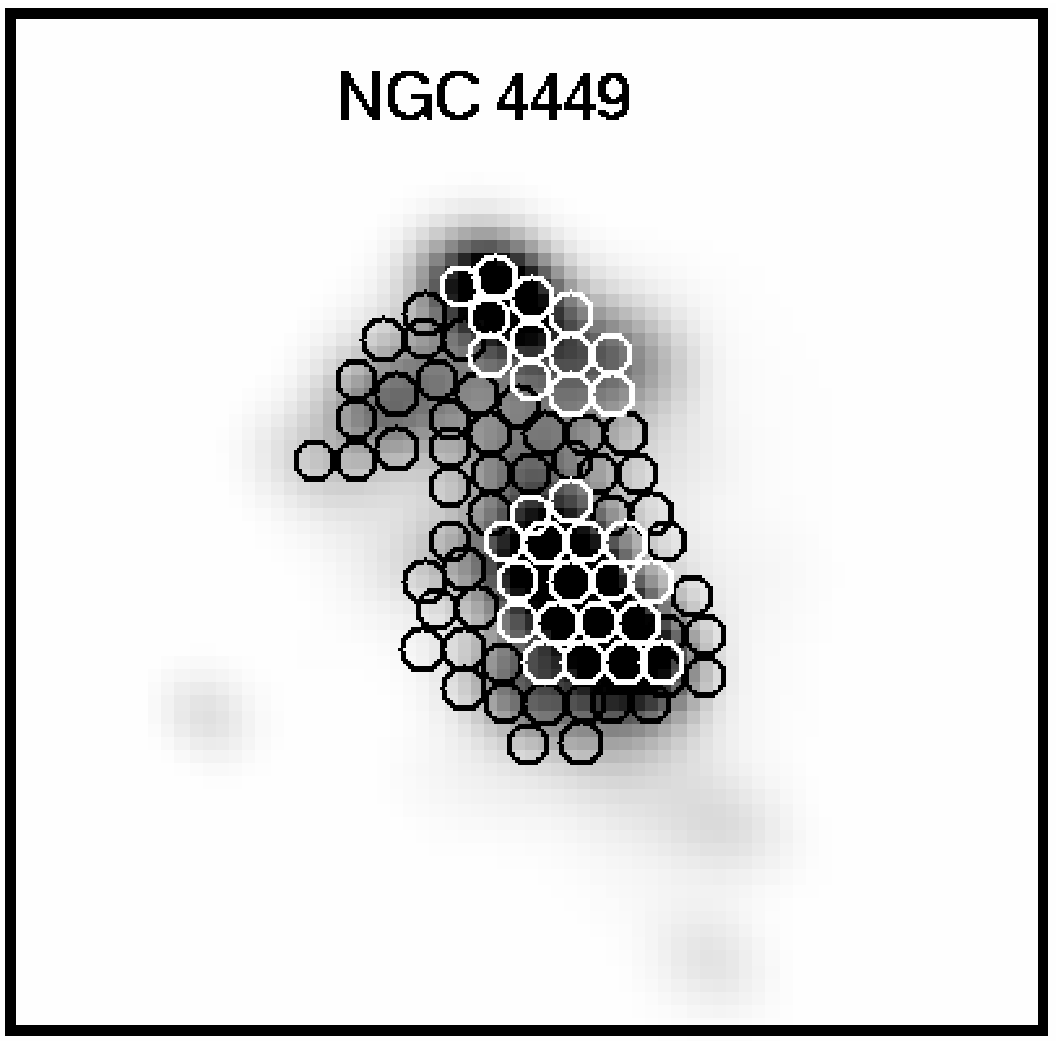}\label{rf2.4449}}
\caption{(a) The radio intensity of NGC 4214 at 325 MHz (solid black symbols)
and 610 MHz (gray symbols) vs. the far infrared intensity at $\lambda70~\mu$m
in units of Jy beam$^{-1}$. The circular and square symbols represents the
bright H{\sc ii} regions and diffuse emission respectively; (b) The radio
intensity of NGC 4449 at 150 MHz vs. the far infrared intensity at
$\lambda70~\mu$m in units of Jy beam$^{-1}$. The circular and square symbols
represents the bright H{\sc ii} regions and diffuse emission respectively; (c)
and (d) The white circles shows the the regions corresponding to the bright
H{\sc ii} regions and the black circles are the diffuse H{\sc ii} regions
overlayed on the H$\alpha$ image of the galaxy (c) NGC 4214 and (d) NGC 4449
after smoothing to 16$''$.}
\end{figure*}

The well known radio-far infrared (FIR) correlation is known to hold good over
five orders of magnitude in luminosity, with a dispersion less than a factor of
2. The radio emission is due to relativistic electrons being accelerated in
ambient galactic magnetic field, while the FIR emission is radiation from dust
heated by ultra-violet (UV) photons from massive ($\gtrsim10~\rm M_\odot$)
short lived ($\sim10^6$ yr) stars.  The two regimes are thought to be connected
by star formation activity \citep{harwi75}. Small-scale amplification of the
magnetic field due to turbulent dynamo action couples the magnetic field ($B$)
and gas density ($\rho_{\rm gas}$) as $B\propto \rho_{\rm gas}^{\kappa}$
\citep[see e.g.,][]{cho00, grove03, chand53} and is believed to produce the
tightness seen in the correlation at local as well as on global scales
\citep[see e.g.,][]{nikla97, dumas11, basu12}.  Here, $\kappa$ is the coupling
index predicted to lie in the range 0.4$-$0.6 by numerical simulations of
different ISM turbulence \citep{fiedl93, grove03, kim01, thomp06, murgi05}.
The slope of the radio--FIR correlation, $b$, defined via $S_{\rm radio} \propto
S_{\rm FIR}^b$ where $S_{\rm radio}$ is the flux density at a radio frequency
and $S_{\rm FIR}$ is the flux density of the FIR emission, is related to the
coupling index $\kappa$ giving rise to the tight radio--FIR correlation
\citep[see e.g.,][]{nikla97, dumas11, basu12}. However, note that this
prescription holds good if both the radio and FIR emission originate from the
same emitting volume. Owing to different propagation lengths of the cosmic ray
electrons (CREs) from the sites of generation, the slope of the correlation in
spatially resolved galaxies is expected to change \citep{basu12, berkh13}.
Here we study spatially resolved radio--FIR correlation at 16$''$ angular
resolution, corresponding to linear scale of $\sim0.23$ and $\sim0.3$ kpc for
the two galaxies NGC 4214 and NGC 4449, respectively. The correlation is
studied using the nonthermal radio emission at 325 and 610 MHz for NGC 4214 and
at 150 MHz for NGC 4449, and the FIR emission at $\lambda70~\mu$m observed
using the {\it Spitzer} space telescope and downloaded from the NED.

Figure~\ref{rf1.4214} shows the nonthermal radio flux density at 325 MHz (solid
black symbols) determined within regions of 16$''$ in size and 610 MHz (gray
symbols) with the FIR flux density at $\lambda70~\mu$m in units of Jy
beam$^{-1}$ for NGC 4214. The circular, square and triangular symbols represent
bright H{\sc ii}, diffuse H{\sc ii} and bridge emission respectively. The
different regions are shown in Figure~\ref{rf2.4214}. The data were fitted with
$S_{\rm radio} = a\times S_{\rm FIR}^b$ using an ordinary least-square bisector
method \citep{isobe90} in the log-log plane at both radio frequencies. The fits
to the data are shown by the dashed lines in Figure~\ref{rf1.4214}.  The
emission at the two wavebands are found to be strongly correlated with
Pearson's correlation coefficient, $r=0.9$ at 325 MHz and $r=0.8$ at 610 MHz.
The slope at 325 MHz is found to be 0.47$\pm$0.04, flatter than that at 610 MHz
where the slope is 0.68$\pm$0.08. 

The magnetic field in NGC 4214 has been estimated to be $\sim30~\mu$G towards
the bright H{\sc ii} regions and $\sim10~\mu$G in the diffuse regions
\citep{kepley2011}.  The magnetic field in such bright HII regions is expected
to be tangled due to turbulence produced by star formation activity and by
propagation of CREs due to streaming instability at Alfv{\'e}n velocity
($v_{\rm A}$, $\sim 50\rm ~km~s^{-1}$). CREs emitting synchrotron radiation at
325 and 610 MHz would propagate $\sim1$ and $\sim$0.6 kpc within synchrotron
cooling timescales. Thus, at 325 MHz, the newly generated CREs and the older
population of CREs are well mixed, giving rise to comparatively higher radio
emission away from the H{\sc ii} regions. This makes the slope of the
radio--FIR correlation flatter at 325 MHz than at 610 MHz for the galaxy NGC
4214. 

Figure~\ref{rf1.4449} shows the plot of nonthermal intensity at 150 MHz with
the $\lambda70~\mu$m intensity for the galaxy NGC 4449. The circles represent
bright H{\sc ii} regions tracing high star formation activity and the squares
represent regions of diffuse H$\alpha$ emission.  Figure~\ref{rf2.4449} marks
the region in the smoothed 16$''$ H$\alpha$ image of the galaxy. The white
circles show the bright star forming regions and the black circles show the
diffuse H$\alpha$ regions.  The slope of the radio--FIR correlation is found to
be 0.75$\pm$0.07 and is significant, with $r = 0.67$.

NGC 4449 shows a steeper slope compared to NGC 4214 for the radio--FIR
correlation when studied at 150 MHz indicating enhanced nonthermal emission in
regions of star-forming sites. Such an effect was also seen for this galaxy at
8.46 GHz \citep{chyzy2000}. The nonthermal spectral index is seen to be flat in
regions of dense star formation, with $\ant\sim-0.5$ (estimated between the 150
and 610 MHz) and the overall nonthermal disc has $\ant = -0.49\pm0.02$.  This likely indicates
that the nonthermal emission in NGC 4449 predominantly originates from freshly
generated CREs which are emitting close to their sites of generation, making
the slope of the radio--FIR correlation steeper.  Since the slope is similar to
that seen in star forming regions in normal galaxies ($\sim 0.8$;
\citet{basu12, basu2014}) it implies that the CREs have propagation length
scales lower than our linear resolution of $\sim 0.3$ kpc.  This allows us to
put upper limits on the age of the CREs assuming they are propagating with an
Alfv{\'e}n velocity of $\sim50\rm ~km~s^{-1}$.    It turns out that the bulk
CREs were produced less than $\sim 6\times 10^6$ ago, due to ongoing star
formation in NGC 4449.  This is supported by the fact that the recent star
formation, as traced by the H$\alpha$ emission of NGC 4449 is about factor of
3--5 higher than that of NGC 4214 \citep{drozdovsky2002, ott2005, hunter1998}.

\section{Conclusions} 

\begin{center}
\begin{table}
\caption{Results}
\label{final.summ}
\scalebox{0.70}{
\begin{tabular}{@{}lllllllll@{}}
\hline 
    Galaxy    & $\alpha_{nth}$  & S$_{th}$ $\%$ & SFR & \multicolumn{3}{c}{b}  \\
              &                 & (610 MHz)  & $10^{-8} M_{\odot} yr^{-1} pc^{-2}$ &  150 MHz & 325 MHz & 610 MHz              \\
\hline
    NGC 4214  & $-0.63(0.04)$ & 22         & $1.97$ & -- & 0.47(0.04) & 0.68(0.08) \\
    NGC 4449  & $-0.49(0.02)$ & 9         & $3.35$  & 0.75(0.07) & --- & --- \\
    \hline \hline
    NGC 4214-I & $-0.32(0.02)$ & 40 & $2.1$ & \\
    NGC 4214-II & $-0.94(0.12)$ & 49 & $2.5$ & \\
\hline
\end{tabular}
}
\\
\end{table}
\end{center}

In this paper, we have presented the low radio frequency images of the WR
galaxies NGC 4214 and NGC 4449 made using the GMRT.  We detect a large radio
halo in NGC 4214 at 325 MHz whose morphology resembles the ultraviolet emission
seen by GALEX with an extent similar to the NIR emission from 2MASS. The
synchrotron spectral index ($\alpha_{nt}$) is $-0.63\pm0.04$.  The thermal
fraction at 610 MHz is about 22\%.  The spectra of two compact star--forming
complexes NGC 4214-I and NGC 4214-II are studied.  We estimate that at 610 MHz
$\sim 40\%$ of emission from NGC 4214-I and $\sim 49\%$ from NGC 4214-II are
thermal in origin.  The non-thermal spectral index of NGC 4214-I is estimated
to be $-0.32\pm0.02$ and of NGC 4214-II is $-0.94\pm0.12$.  The combination of
compact star-forming regions and presence of an extended halo make this dwarf
galaxy similar to a normal star-forming galaxy.   This galaxy follows the
radio-FIR relation.  We find a significant correlation between the local 325
MHz and 610 MHz emission and the 70 $\mu$m emission for the entire galaxy, with
slopes of 0.47 and 0.68 respectively, which is within the range expected from
simulations of turbulence in the interstellar medium.  

We detect a large radio halo around NGC 4449 at 150 MHz. 
NGC 4449 is estimated to be five time more luminous than NGC 4214, indicating
vigorous star formation in the former.  Both the galaxies are in a group
environment and hence their star formation properties are  likely to be tidally
influenced by the other group members.  
Separating the non-thermal from the thermal emission which is obtained from the
H$\alpha$ map of NGC 4449 results in $\alpha_{nt}$ of $-0.49\pm0.02$.  For NGC
4449 we estimate a thermal fraction at 610 MHz of $\sim 9\%$ and find that the
150 MHz emission is well correlated with the 70 $\mu$m emission on small
scales, with a slope of 0.75. The non-thermal spectral index of both galaxies
is flatter than estimated for normal disk galaxies.  

\section{Acknowledgements} {\small We thank the staff of GMRT who made these
observations possible. GMRT is run by the National Centre for Radio
Astrophysics (NCRA) of the Tata institute of Fundamental Physics. This research has
made use of the NASA/IPAC Extragalactic Database (NED) which is operated by the
Jet Propulsion Laboratory, California Institute of Technology, under contract
with the National Aeronautics and Space Administration. This research has also
made use of the GALEX and {\it Spitzer Space Telescope} which are NASA mission
managed by the Jet Propulsion Laboratory. We also like to acknowledge the site
http://skyview.gsfc.nasa.gov/.  SA acknowledges an INSA Senior scientist fellowship.
SS gratefully acknowledges the support of a
research grant SR/S2/HEP-08/2008 by the Department of Science $\&$ Technology
(DST), Govt. of India and support from NCRA. }

\bibliographystyle{mn2e} \footnotesize \bibliography{bibliography}

\label{lastpage}
\end{document}